\documentclass{aa}  

\usepackage{graphicx}
\usepackage{txfonts}
\usepackage{graphicx}
\usepackage{caption}
\usepackage{longtable}
\usepackage{multicol}
\usepackage{color}
\usepackage{lscape}
\usepackage{ulem}
\usepackage[colorlinks, citecolor=blue,urlcolor=blue, linkcolor=blue]{hyperref}
\begin{document}

\title{Probing fragmentation and velocity sub-structure in the massive NGC~6334 filament with ALMA}


   \author{Y. Shimajiri\inst{1,2,3}, Ph. Andr$\acute{\rm e}$\inst{1}, E. Ntormousi\inst{4,5}, A. Men'shchikov\inst{1}, D. Arzoumanian\inst{6,7}, and P. Palmeirim\inst{7} 
          }

   \institute{
\inst{1} Laboratoire d'Astrophysique (AIM), CEA/DRF, CNRS, Universit\'{e} Paris-Saclay, Universit\'{e} Paris Diderot, Sorbonne Paris Cit\'{e}, 91191 Gif-sur-Yvette, France\\
              \email{Yoshito.Shimajiri@cea.fr, pandre@cea.fr}\\
\inst{2} Department of Physics and Astronomy, Graduate School of Science and Engineering, Kagoshima University, 1-21-35 Korimoto, Kagoshima,
Kagoshima 890-0065, Japan\\
\inst{3} National Astronomical Observatory of Japan, Osawa 2-21-1, Mitaka, Tokyo 181-8588, Japan\\
\inst{4} Foundation for Research and Technology (FORTH), Nikolaou Plastira 100, Vassilika Vouton GR - 711 10, Heraklion, Greece\\
\inst{5} Department of Physics and ITCP,  University of Crete, 71003 Heraklion, Greece\\
\inst{6} Department of Physics, Graduate School of Science, Nagoya University Nagoya 464-8602, Japan\\
\inst{7} Instituto de Astrofisica e Ciencias do Espaco, Universidade do Porto, CAUP, Rua das Estrelas, PT4150-762 Porto, Portugal\\
             }
   \date{Received 12 April 2019 / Accepted 3 October 2019 }

 
 \abstract
 {{\it Herschel} imaging surveys of galactic interstellar clouds support a paradigm for low-mass star formation 
 in which dense molecular filaments play a crucial role. The detailed fragmentation properties of star-forming filaments remain poorly understood, however, 
 and the validity of the filament paradigm in the intermediate- to high-mass regime is still unclear. }
 {Here, following up on an earlier 350 $\mu$m dust continuum study with the ArT$\acute{\rm e}$MiS camera on the 
 APEX telescope, we investigate the detailed density and velocity structure of the main filament in the high-mass star-forming region NGC~6334.
 }
 {We conducted ALMA Band\,3 observations in the 3.1 mm continuum and of the N$_2$H$^+$(1-0), HC$_5$N(36-35), HNC(1-0), HC$_3$N(10-9), CH$_3$CCH(6-5), and H$_2$CS(3-2) lines  
 at an angular resolution of $\sim$3$\arcsec$, corresponding to 0.025 pc at a distance of 1.7 kpc.}
 {The NGC~6334 filament was detected in both the 3.1 mm continuum and the N$_2$H$^+$, HC$_3$N, HC$_5$N, CH$_3$CCH, and H$_2$CS lines with ALMA. 
We identified twenty-six compact ($< 0.03\,$pc)
dense cores at 3.1 mm and five velocity-coherent fiber-like features 
in N$_2$H$^+$ within the main filament. The typical length ($\sim \,$0.5~pc) 
of,  and velocity difference ($\sim \,$0.8 km s$^{-1}$) between, 
the fiber-like features 
of the NGC~6334 filament are reminiscent of the properties for the fibers of the low-mass star-forming filament B211/B213 in the Taurus cloud.
Only two or three of the five velocity-coherent features are well aligned with the 
NGC 6334 filament and may represent genuine, fiber sub-structures; 
the other two 
features may trace accretion flows onto the main filament.
The mass distribution of the ALMA 3.1 mm continuum cores has a peak at {$\sim 10\, M_\odot $}, which is an order of magnitude higher 
than the peak of the prestellar core mass function in nearby, low-mass star-forming clouds.
The cores can be divided into seven groups, 
closely associated with dense clumps seen in the ArT$\acute{\rm e}$MiS 350 $\mu$m data. 
The projected separation between ALMA dense cores (0.03--0.1 pc) and the projected spacing between ArT$\acute{\rm e}$MiS clumps (0.2--0.3 pc)
are roughly consistent with the effective Jeans length ($0.08\pm0.03\, $pc) 
in the filament 
and a physical scale of about four times the filament width, respectively, 
if the inclination angle of the filament to line of sight is $\sim 30^\circ $.
These two distinct separation scales are suggestive of a bimodal fragmentation process in the filament.
}
{Despite being one order of magnitude denser and more massive than the Taurus B211/B213 filament, the NGC~6334 filament has a density and velocity structure 
that is qualitatively very similar. 
The main difference is that the dense cores embedded in the NGC~6334 filament appear to be an order of magnitude denser 
and more massive than the cores in the Taurus filament. This suggests that dense molecular filaments may evolve and fragment 
in a similar manner in low- and high-mass star-forming regions, and that 
the filament paradigm may hold in the intermediate-mass  (if not high-mass) star formation regime.
}

   \titlerunning{Structure of the NGC~6334 filament}
   \authorrunning{Shimajiri et al.}
\maketitle
%

\begin{table*}
\centering
\caption{Parameters for the ALMA observations \label{table:alma_obs}}
\begin{tabular}{lcccc}
\hline
                                                   &  Rest freq.     &  $\theta_{\rm beam}$(PA) & rms &  dV \\
                                                   &   GHz                  & $\arcsec \times \arcsec$ ($^{\circ}$) & mJy/beam & km/s \\
\hline
3.1 mm continuum (12m+7m)                                 &  --------------  &  2.99 $\times$ 1.97 (82.6)  &  0.2  & ---- \\
\textcolor{black}{3.1 mm continuum (12m)}                               &  --------------  &  2.80 $\times$ 1.84 (83.4)  &  0.14  & ---- \\
HNC (1--0,$E_{\rm u}$=4.35 K)                                 &  90.663572           &  3.82 $\times$ 2.48 (85.9)  &   0.3  & 60  \\
HC$_3$N (10--9,$E_{\rm u}$=24.015 K)                       & 90.978989            & 3.81 $\times$ 2.47 (85.9) &   0.27  & 60  \\
N$_2$H$^+$ (1--0,$E_{\rm u}$=4.472 K)                   & 93.176265            & 3.78 $\times$ 2.40 (84.1) & 4.4  & 0.2  \\
HC$_5$N (36--35,$E_{\rm u}$=80.504 K)                      & 93.188126           & 3.77 $\times$ 2.39 (84.1) & 5.4 & 0.2  \\
CH$_3$CCH (6$_0$--5$_0$,$E_{\rm u}$=17.225 K)      & 102.54798          &  3.75 $\times$ 2.36 (79.1) & 0.36 & 60 \\
H$_2$CS (3$_{1,2}$-2$_{1,1}$,$E_{\rm u}$=23.213 K)   & 104.61699          & 3.40 $\times$ 2.17 (84.3) & 0.22  & 60  \\
\hline
\end{tabular}
\end{table*}

\section{Introduction}\label{intro}

{\it Herschel} imaging observations of galactic molecular clouds 
reveal an omnipresence of filamentary structures 
and suggest that filaments dominate the mass budget of the dense molecular gas 
where stars form \citep{Andre10, Molinari10, Hill11, Schisano14, Konyves15}. 
At least in the nearby clouds of the Gould Belt, detailed studies of the radial column density profiles 
have found a common inner filament width of $\sim 0.1\,$pc, with a dispersion of a factor $\la 2$,  
when averaged over the filament crests \citep{Arzoumanian11,Arzoumanian18, Palmeirim13, Koch15}. 
Furthermore, most of the prestellar cores identified with {\it Herschel} are found to be embedded 
within such filamentary structures, showing that dense molecular filaments are the main sites of solar-type star formation \citep[e.g.,][]{Konyves15,Konyves19,Marsh16}. 
Overall, the {\it Herschel} findings in nearby clouds 
support a filament paradigm for solar-type star formation in two main steps (cf. \citealp{Andre14}, \citeyear{Andre17}; \citealp{Inutsuka15}):
first, multiple large-scale compressions of interstellar material in supersonic flows generate a pervasive web of $\sim$0.1-pc wide 
filaments in the cold interstellar medium (ISM); second, the densest filaments fragment into prestellar cores by gravitational instability 
near or above the critical mass per unit length of nearly isothermal gas cylinders, $M_{\rm line, crit} = 2\, c_{\rm s}^2/G \sim 16\, M_\odot$/pc, 
where $c_{\rm s} \sim 0.2$~km s$^{-1}$ is the isothermal sound speed for molecular gas at $T \sim 10$~K. 

Since molecular filaments are also known to be present in other galaxies \citep[cf.][]{Fukui15}, this paradigm 
may potentially have implications for star formation on galaxy-wide scales. 
The star formation efficiency in dense molecular gas is indeed observed to be remarkably uniform 
over a wide range of scales from pc-scale filaments and clumps to entire galactic disks \citep{Gao04,Lada10, Lada12, Shimajiri17, Zhang19}, 
with possible deviations in extreme environments, such as the central molecular zone \citep{Longmore13}.  
Assuming that all star-forming filaments have similar inner widths, as seems to be the case in nearby clouds \citep{Arzoumanian18}, 
it is argued that the microphysics of filament fragmentation into prestellar cores may ultimately be responsible 
for this quasi-universal star formation efficiency \citep{Andre14,Shimajiri17}.

The validity and details of the two-step filament paradigm remain controversial, however, especially beyond the Gould Belt and in high-mass star-forming clouds. In particular, an alternative scenario is proposed based on the notion of global hierarchical cloud collapse \citep[e.g][]{Vazquez19}, which is especially attractive in the case of high-mass star formation to account for the structure of strongly self-gravitating hub-filament systems, where a massive cluster-forming hub is observed at the center of a converging network of filaments \citep{Myers09,Peretto13}. In this alternative scenario, most if not all filaments would be generated by gravitational effects, as opposed to large-scale compression, and would represent accretion flows onto dense hubs.

The evolution and detailed fragmentation manner of star-forming filaments are also poorly understood. 
In particular, the mere existence of massive, $\sim$0.1-pc wide, filaments with masses per unit length $M_{\rm line}$
that exceed the critical line mass\footnote{The critical line mass is $M_{\rm line,crit} = 2\,c_{\rm s}^2$/G, 
where $c_{\rm s}$ is the isothermal sound speed, corresponding to  $M_{\rm line,crit} \approx  33\, M_{\odot}$/pc for a molecular gas temperature $T$ = 20 K.} of an isothermal filament $M_{\rm line, crit}$ by one to two orders of magnitude is a paradox. Indeed, such filaments 
may be expected to undergo rapid radial contraction into spindles before any fragmentation into prestellar cores \citep[cf.][]{Inutsuka97}. 
A possible solution for this paradox is that massive filaments accrete background cloud material while evolving \citep[cf.][]{Palmeirim13,Shimajiri18}, 
and that this accretion process drives magneto-hydrodynamic (MHD) waves, generating sub-structure within dense filaments, and 
leading to a dynamical equilibrium \citep{Hennebelle13}, where $M_{\rm line}$ approaches 
the virial mass per unit length $M_{\rm line, vir} = 2\, \sigma _{\rm 1D}^2/G $ \citep[][where $\sigma _{\rm 1D}$ is the one-dimensional velocity dispersion]{Fiege00}. 
The detection through molecular line observations of velocity-coherent fiber-like sub-structure 
within several nearby supercritical filaments \citep{Hacar13, Hacar18} may possibly be the manifestation of such a process.

The physical origin of observed fibers is nevertheless still under debate \citep{Tafalla15, Smith14,Clarke17,Clarke18}, and 
it is not yet clear whether dense molecular filaments in massive star-forming regions have similar characteristics to those observed in nearby clouds. 
Our recent APEX/ArT$\acute{\rm e}$MiS 350 $\mu$m study of the 
massive star-forming complex NGC~6334 
showed that the main filament of the complex has an observed line mass of $\sim$1000 $M_{\odot}$/pc, 
consistent within uncertainties with the estimated virial mass per unit length $\sim$500 $M_{\odot}$/pc, 
and an inner width $\sim 0.15\pm 0.05\,$pc all along its length \citep{Andre16}, 
within $\sim$$50\% $ 
of the typical inner filament width observed with {\it Herschel} in nearby clouds \citep{Arzoumanian18}.
NGC~6334 is a very active star-forming region at a distance $d \sim 1.7\,$kpc, 
which contains 150 OB stars and more than 2000 young stellar objects \citep{Persi08, Russeil13, Willis13,Tige17},  
for a total gas mass of $\sim$2.2$\times$10$^5$ $M_{\odot}$.
Here, thanks to the high angular resolution and sensitivity of Atacama Large Millimeter Array (ALMA) data at 3 mm, 
we investigate the density and velocity sub-structure of the massive NGC~6334 filament and compare the results 
with those obtained for nearby, lower-mass star-forming filaments. 

The paper is organized as follows. 
In Sect. \ref{ALMA_obs}, we describe our ALMA observations of NGC~6334 in both the 
N$_2$H$^+$(1--0), HC$_3$N(10--9), HC$_5$N(36--35), CH$_3$CCH(6$_0$--5$_0$), H$_2$CS(3$_{1,2}$--2$_{1,1}$) lines 
and  the 3.1 mm continuum. 
In Sect. \ref{Sect:Results}, we show the spatial distributions of the detected molecular lines and continuum emission. 
We also analyze the velocity features observed in N$_2$H$^+$(1--0) emission and extract dense structures such as compact cores and fiber-like components. 
In Sect. \ref{Sect:Discussion}, we discuss the evidence of a bimodal fragmentation pattern in the NGC~6334 filament, 
emphasize the presence of both unusually massive cores and 
fiber-like velocity-coherent components in the filament, 
and comment on the possible origin of these multiple velocity components. 
Our conclusions are summarized in Sect.~\ref{Sect:Conclusions}.

\begin{figure*}
\centering
\includegraphics[width=170mm, angle=0]{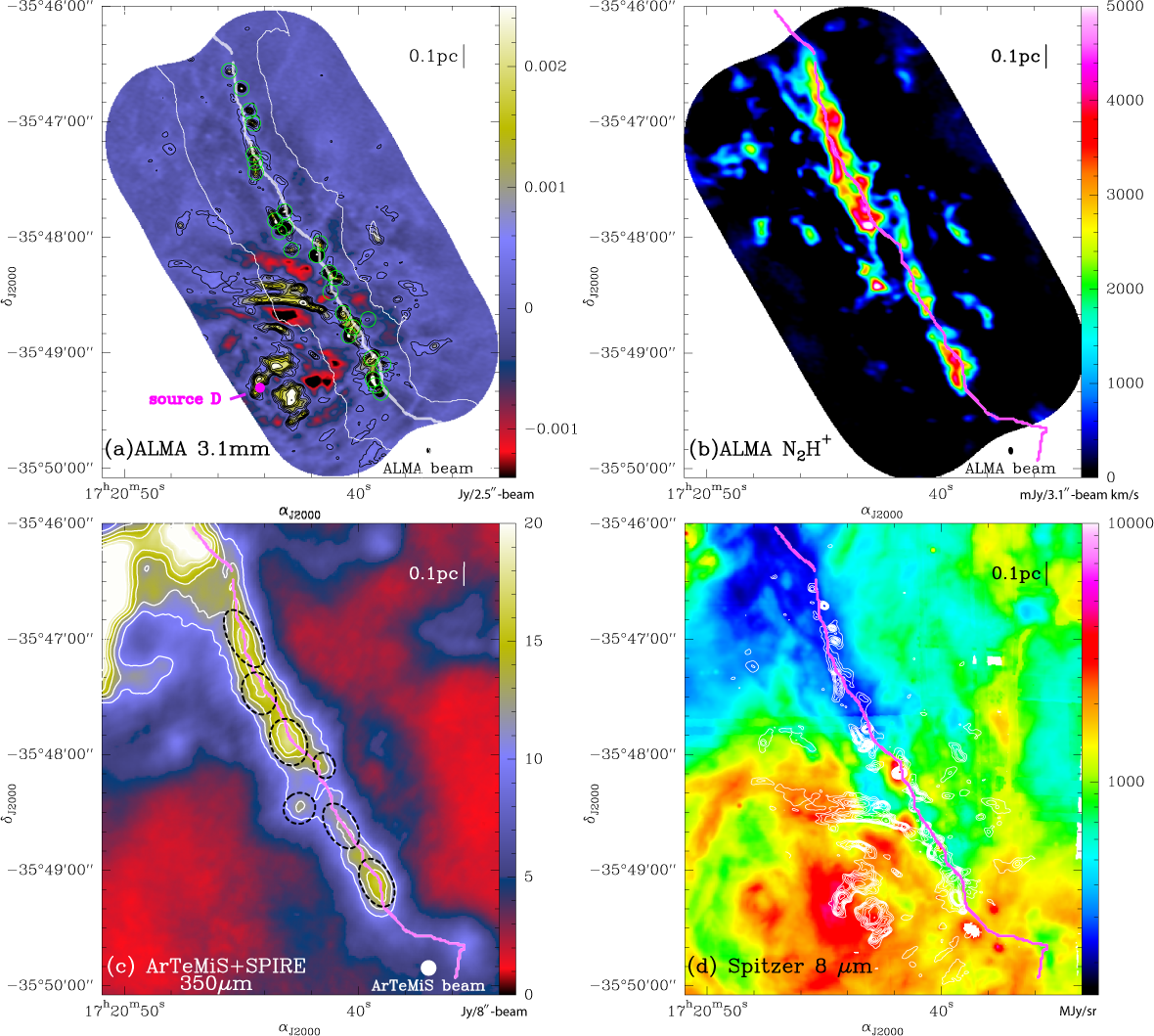}
\caption{Maps of the NGC~6334 main filament region in 3.1 mm continuum emission with ALMA 12m (panel $a$), N$_2$H$^+$(1--0) integrated intensity with ALMA 12m+7m (panel $b$), 350 $\mu$m dust continuum emission with ArT$\acute{\rm e}$MiS (panel $c$,  from \citealp{Andre16}), and 8 $\mu$m dust continuum emission with {\it Spitzer} (panel $d$). In panel ($a$), the magenta filled circle indicates the position of a radio continuum compact HII region (source D in \citealp{Rodriguez82}), and green open circles indicate the positions of the compact 3.1 mm continuum sources identified with \textsl{getsources} (see Sect.~\ref{sect:getsources}). In panel ($a$), the solid white contour marks the footprint of the main filament, defined as the intersection of the area within $\pm$30$\arcsec$ 
from the filament crest and the interior of the 5~Jy\,beam$^{-1}$ contour in the ArT$\acute{\rm e}$MiS 350 $\mu$m map.
In panels ($a$), ($d$), contours of 3.1 mm continuum emission are overlaid with levels of 2, 4, 6, 8, 10, 15, 20, 25, 30, 35, and 
40$\sigma$, where 1$\sigma$ = 0.2 mJy/ALMA-beam. 
In panel ($b)$, the N$_2$H$^+$ intensity was integrated over the velocity range from $-$11.4 km s$^{-1}$ to $+$16.8 km s$^{-1}$, including all hyperfine
components.
In panel ($c$), contours of 350 $\mu$m dust continuum emission are shown with levels of 10, 12, 14, 16, 18, and 20~Jy/8\arcsec -beam, and dashed ellipses indicate the positions of the ArT$\acute{\rm e}$MiS dense clumps identified with GAUSSCLUMPS (see Sect.~\ref{two_levels}). In all panels, the crest of the NGC 6334 main filament as traced with the DisPerSE algorithm in the ArT$\acute{\rm e}$MiS 350 $\mu$m map \citep{Andre16} is shown as a solid curve.
}
\label{figs}
\end{figure*}

\section{ALMA observations of the NGC 6334 filament}\label{ALMA_obs}

We carried out ALMA Cycle 3 observations in Band 3 
toward NGC 6334 with both the 12m antennas (C36-2 configuration) and the Atacama Compact
Array (ACA) 7m antennas, as part of project 2015.1.01404.S (PI: Ph. Andr$\acute{\rm e}$). 
We imaged the main filament 
in the NGC 6334 region using a 17-pointing mosaic with the 12m array and a 8-pointing mosaic with the ACA array. 
The N$_2$H$^+$(1--0) line was observed in narrow-band mode at a frequency resolution of 61.035 kHz, corresponding to $\sim\,$0.2 km s$^{-1}$. 
The HC$_5$N ($J$=36--35) line was included in the same narrow-band, high spectral resolution window. 
The 3.1 mm continuum emission was observed using three wide bands, each covering a bandwidth of 1875.0 MHz. 
The HNC (1--0), HC$_3$N (10--9), CH$_3$CCH (6$_0$--5$_0$), and H$_2$CS (3$_{1,2}$-2$_{1,1}$) lines 
were included in these wide bands and observed at a frequency resolution of 31.250 MHz, corresponding 
to $\sim\,$100 km s$^{-1}$.  
The 12m-array observations were carried out on 23, 24, 26 January 2016 with  42, 46, and 37 antennas, respectively,
and projected baseline lengths ranging from 11.7 k$\lambda$ to 326.8 k$\lambda$.
The ACA observations were carried out between 16 March 2016 and 8 July 2016 with 7-9 antennas and projected baseline lengths ranging from 4.5 k$\lambda$ to 48.2 k$\lambda$.
In the calibration process, we made additional flagging of a few antennas that had too
low gain or showed large amplitude dispersion in time. 
The bandpass calibration was achieved by observing the quasar J1617-5848 with the 12m array and the three quasars J1924-2914, J1517-2422, and J1427-4206 with the ACA array. 
The complex gain calibration was carried out using the quasar J1713-3418 with the 12m array and the quasar J1717-3342 with ACA. 
Absolute flux calibration was achieved by observing two solar system objects (Callisto and Titan) with the 12m array and planets (Mars and Neptune) with ACA.
Calibration and data reduction were performed 
with the Common Astronomy Software Application (CASA) package (version 4.5.3 for 12m data calibration and imaging, version 4.6.0 for 7m data calibration). 
For imaging, we adopted the Briggs weighting scheme with a robust parameter 
of 0.5, as a good compromise between maximizing angular resolution and sensitivity to extended structures.
The resulting synthesized beam sizes ($\sim$3$\arcsec$, corresponding to 0.025 pc)
and rms noise levels for each line and the continuum are summarized in Table \ref{table:alma_obs}.

With a minimum projected baseline length of 4.5 k$\lambda$, our ALMA$+$ACA observations 
are estimated to be sensitive to angular scales up to $\sim$37$\arcsec$ (corresponding to $\sim$0.3 pc) at the 10\% level \citep{Wilner94}. 
For comparison, we expect the transverse size of any filament sub-structures to be smaller than 
the $\sim$0.15\,pc inner width of the NGC~6334 filament \citep{Andre16}.
Likewise with a minimum projected baseline of  11.7 k$\lambda$, our 12m-only observations 
are sensitive to angular scales up to $\sim$14$\arcsec$ (corresponding to $\sim$0.1 pc) at the 10\% level. 
For comparison, the scales of individual dense cores are at most $\sim$0.1 pc \citep[][]{Konyves15}.
Moreover, simulations of  ALMA observations performed with CASA,  
using the same set of baselines as the real ALMA$+$ACA data, show that interferometric-filtering effects 
do not generate spurious sub-structures within the filament (see Fig. \ref{fig_alma_sim} in Appendix~A).

\section{Results and analysis}\label{Sect:Results}

In this section, we show the results of our ALMA 3.1 mm continuum and N$_2$H$^+$ (1--0) line observations, 
and then extract compact 3.1 mm continuum sources and fiber-like velocity-coherent structures from the ALMA data.

\subsection{3.1 mm continuum emission} \label{result:cont}
Our ALMA 3.1 mm continuum map of the NGC 6334 filament region is shown in panel $a$ of Fig.~\ref{figs}  (color scale and contours). 
It is also overlaid as contours 
on a {\it Spitzer} 8\,$\mu$m emission map in panel $d$ 
of  Fig.~\ref{figs}.
Hereafter, we call the map resulting from the combination of APEX/ArT$\acute{\rm e}$MiS 350 $\mu$m \citep[][]{Andre16}
and {\it Herschel}/SPIRE 350 $\mu$m \citep[][]{Russeil13} data the ArT$\acute{\rm e}$MiS 350 $\mu$m map. 
The counterpart of the dense filament seen in the ArT$\acute{\rm e}$MiS 350 $\mu$m map can be 
recognized in the northern part of the ALMA 3.1 mm continuum map.
In the southern part of the field, the filament is not clearly detected in the ALMA 3.1 mm continuum map, 
but a shell-like structure can be seen. 
Conversely, the shell-like structure is not seen in the ArT$\acute{\rm e}$MiS 350 $\mu$m map. 
A compact HII region associated with a 4.9 GHz radio continuum source labeled source D in \citet{Rodriguez82} lies close 
to the contours of this shell-like structure.
The shell-like structure 
in the ALMA 3.1 mm continuum map also 
coincides with bright, extended mid-infrared emission detected at 8\,$\mu$m  with {\it Spitzer} 
as shown in Fig.~\ref{figs}$d$.
We conclude that the 3.1 mm continuum emission in the shell-like structure is most likely dominated 
by free-free emission from the compact HII region around the luminous embedded 
star associated with source D.

\subsection{Molecular lines}\label{sect:molecular_lines}

\subsubsection{Spatial distribution of the N$_2$H$^+$(1--0) emission}\label{sect:molecular_lines_dist1}

Figure~\ref{figs}$b$ shows the integrated intensity map of the N$_2$H$^+$(1--0) line emission around the NGC~6334 main filament. 
It can be seen that 
the prominent dusty filament in the  ArT$\acute{\rm e}$MiS 350 $\mu$m map (Fig.~\ref{figs}$c$) 
is very well traced by the ALMA N$_2$H$^+$(1--0) data.
In addition, a few N$_2$H$^+$ blobs can be recognized outside the main filament. 

\subsubsection{Spatial distributions of other molecular line tracers}\label{sect:molecular_lines_dist2}

The ALMA maps obtained in all other molecular line tracers, except HNC(1--0), also show the NGC 6334 filament 
(see Fig.~\ref{figs_otherlines}$a$--$e$ in Appendix~\ref{Sect:Complementary}).
The map obtained in HNC(1--0) differs from the other maps in that it shows a rather scattered distribution of discrete blobs.
Due to a lower effective excitation density \citep[cf.][]{Shirley15}, the emission from 
the HNC(1--0) transition may be more extended than the emission in the other dense gas tracers observed here 
and may be resolved out due to interferometric filtering effects, even with ACA. 
The maps obtained in CH$_3$CCH(6$_0$--5$_0$), H$_2$CS(3$_{1,2}$--2$_{1,1}$), HC$_3$N(10--9), and HC$_5$N(36--35) emission 
appear to trace the same filamentary structure as detected in N$_2$H$^+$(1--0). 
The HC$_5$N(36--35) line has a high upper state energy of 80.504 K, implying that some of the gas in the 
main filament has a high temperature and/or density. 
Indeed, \citet{Andre16} estimated the average column density 
and average volume density of the entire NGC 6334 filament to be 1-2$\times$10$^{23}$ cm$^{-2}$ and 2.2$\times$10$^{5}$ cm$^{-3}$, respectively. 
This exceeds the (column) density values observed by \citet{Palmeirim13} for the low-mass star-forming filament B211/B213 in Taurus 
($\sim$1.4$\times$10$^{22}$ cm$^{-2}$ and 4.5$\times$10$^{4}$ cm$^{-3}$) 
by an order of magnitude.

\subsubsection{Observed line profiles}\label{Sect:velo_structure}

\begin{figure}
\centering
\includegraphics[width=100mm, angle=0]{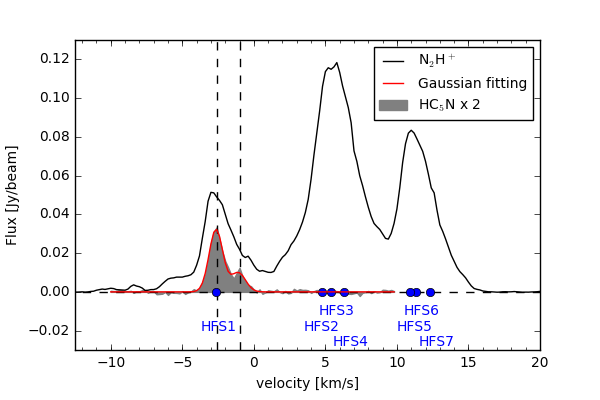}
\caption{(black) N$_2$H$^+$(1--0) and (gray) HC$_5$N(36--35) spectra averaged over pixels where the emission was detected above 5 $\sigma$. 
The red curve shows a two-component Gaussian fit to the HC$_5$N line profile.
The blue circles indicate the peak velocity positions of each N$_2$H$^+$(1--0) HFS component 
expected from the HC$_5$N(36--35) systemic velocity of $-$2.6 km s$^{-1}$. 
The vertical dashed lines mark the peaks of the two HC$_5$N velocity components. 
}
\label{figs:hc5n_n2h+_spectrum}
\end{figure}

\begin{figure*}
\centering
\includegraphics[width=1.0\textwidth, angle=0]{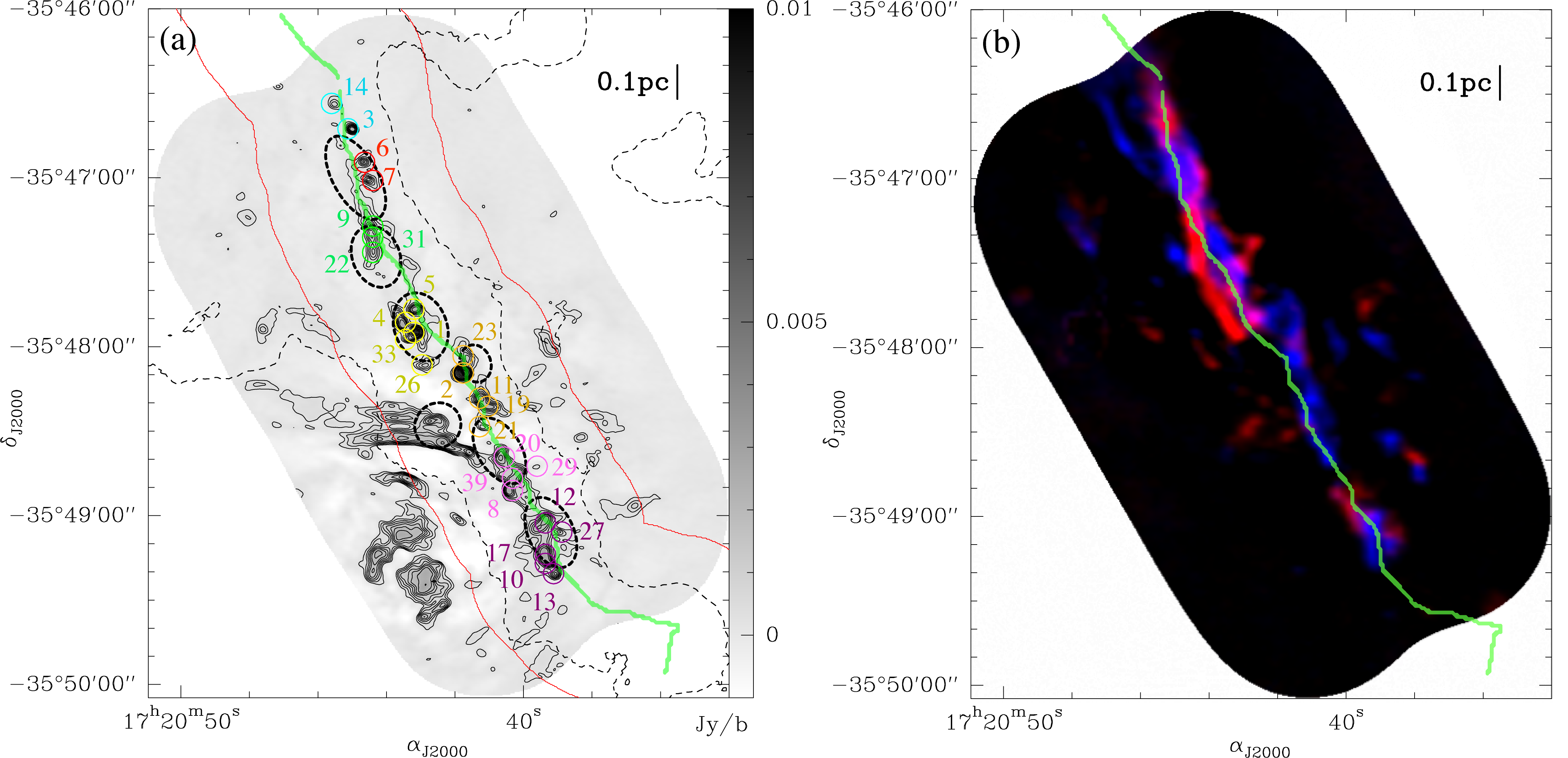}
\caption{({\it a}) Spatial distribution of the 26 ALMA compact cores (magenta, pink, orange, yellow, green, red, green, and light blue circles) and 7 ArT$\acute{\rm e}$MiS clumps (larger black dashed ellipses-- see Sect.~\ref{two_levels}) embedded in the NGC~6334 main filament, overlaid on the ALMA 3.1~mm continuum image (grayscale and same solid contours as in Fig. \ref{figs}). The black dashed contour shows an ArT$\acute{\rm e}$MiS 350 $\mu$m continuum emission level of 5 Jy/8$\arcsec$-beam. 
The green curve marks the crest of the dusty filament traced with DisPerSE in the ArT$\acute{\rm e}$MiS 350 $\mu$m continuum map \citep[][see Fig.~\ref{figs}c]{Andre16}. 
The red curves mark the area within $\pm$30$\arcsec$ (or $\pm$0.25~pc)  from this filament crest.
({\it b}) Spatial distribution of the N$_2$H$^+$(1--0) velocity-coherent fiber-like sub-structures F-1 (blue, integrated over the velocity range $-$3 km s$^{-1}$ to $-$2.4 km s$^{-1}$) and F-2 (red, integrated over the velocity range $-$2.0 km s$^{-1}$ to $-$1.2 km s$^{-1}$) 
in the NGC~6334 filament. In panel {\it a}, the seven groups of ALMA 3.1 mm compact sources found by the nearest neighbor separation algorithm (see Sect. \ref{two_levels} and Fig.~\ref{figs:NNS_result}) are indicated by small circles of different colors. 
The green curve is the same as in panel ({\it a}).}
\label{figs:DNA_fiber}
\label{figs:ArTeMiS_ALMA_core_association}
\end{figure*}

The N$_2$H$^+$(1--0) and HC$_5$N(36--35) line observations were obtained in narrow-band mode 
with a velocity resolution of 0.2 km s$^{-1}$, which allows us to investigate the velocity structure
of the NGC~6334 
filament. 

Here, we discuss the velocity profiles obtained in N$_2$H$^+$(1--0) and HC$_5$N(36--35). 

Figure~\ref{figs:hc5n_n2h+_spectrum} shows the N$_2$H$^+$(1--0) and HC$_5$N(36--35) spectra averaged over the pixels 
where significant emission was detected above the 5$\sigma$ level. Two peaks at $V_{\rm sys}$=$-$2.6 and $V_{\rm sys}$=$-$1.0 km s$^{-1}$ can be recognized in the HC$_5$N(36--35) spectrum. 
We set the velocity scale of the N$_2$H$^+$ (1--0) spectrum using the rest frequency of the isolated component of the N$_2$H$^+$ ($v$=0, $J$=1--0, $F_1$=0--1, $F$=1--2, 93.176265 GHz) hyperfine structure (HFS) multiplet as a reference.
Hereafter, we call this isolated N$_2$H$^+$ component "HFS 1" and the other components "HFS 2-7" (see Fig.~\ref{figs:hc5n_n2h+_spectrum}). 
The peak velocity of HFS 1 ($\sim$ $-$2.6 km s$^{-1}$) is consistent with the peak velocity of the HC$_5$N(36--35) emission. 
In addition, the N$_2$H$^+$ HFS 1 line profile exhibits an emission wing at highly blueshifted velocities (up to $-$12 km s$^{-1}$, see Fig. \ref{figs:hc5n_n2h+_spectrum}). 
This blueshifted emission does not appear to be associated with the NGC 6334 filament itself since it is mainly detected  
outside the main filamentary structure (see Fig. \ref{figs:n2h+_blue}).
At positions where the blueshifted emission between $-$12  to $-$6 km s$^{-1}$ overlaps with 
the main velocity component at $\sim$ $-$2.6 km s$^{-1}$ in the plane of the sky, the HFS 2-7 counterparts of the blueshifted emission 
can be confused with HFS~1 emission from the  $-$2.6 km s$^{-1}$ component 
since the velocity offsets of the HFS 2-7 components range from 7.3 km s$^{-1}$ to 14.9 km s$^{-1}$ 
with respect to the velocity of HFS~1 (see Appendix~\ref{Sect:Contamination}).

\subsection{Extraction of compact continuum sources and "fiber-like" velocity structure from the data}\label{Sect:Filament_ID}

To extract compact continuum sources from the ALMA/ACA 3.1 mm continuum map
and fiber-like velocity-coherent features from the ALMA/ACA N$_2$H$^+$ (1--0)  data cube, 
we made use of the \textsl{getsources}, \textsl{getfilaments}, and \textsl{getimages} algorithms \citep[][and in prep.]{Menshchikov12, Menshchikov13,Menshchikov17}.


\begin{table*} 
\tiny 
\centering 
\caption{Results of compact source extractions on the ALMA 12m-only 3.1 mm  continuum map \label{table:cont_getsources_12m_only}}  
\scalebox{0.8}{ 
\begin{tabular}{lcccccccccccc} 
\hline 
ID$^{1}$  & RA & DEC & $R_{\rm major} \times R_{\rm minor}$ & PA      &  S$_{\rm 3.1mm}^{\rm peak}$ & S/N   & $S_{\rm 3.1mm}^{\rm tot}$ & $M_{\rm tot}$ & $\overline{n_{\rm H_2}}$ & Velocity channel at which   & Association with   \\ 
&  &  &     &   & & & &  & & source is detected in N$_2$H$^+$  &  N$_2$H$^+$ features  \\ 
     & [hms] & [dms] & [$\arcsec \times \arcsec$]$^{3,4}$ & [deg]$^{3}$ &   [mJy/beam]$^{3}$ & [$\sigma$] & [mJy]$^{3}$ & [$M_{\odot}$]$^{5}$  & [$\times$10$^6$ cm$^{-3}$]$^{6}$  & [km s$^{-1}$]  &    \\ 
\hline 
{\bf 1 }  & 17:20:43.13 & -35:47:54.9 & 2.9$\times$2.3 (0.02 pc $\times$ 0.02 pc) & 1 & 5.1$\pm$0.2 & 36.4 & 11.4 & 31.5 & 11.1 & [-2.2,-1.8] & F-2 \\  
2  & 17:20:41.78 & -35:48:9.2 & 2.8$\times$1.7 (0.02 pc $\times$ 0.01 pc) & 10 & 6.9$\pm$0.2 & 49.2 & 13.5 & 13.8 & 8.1 &  & No N$_2$H$^{+}$  \\  
3 & 17:20:45.00 & -35:46:42.7 & 1.7$\times$0.8 (0.01 pc $\times$ 0.01 pc) & -33 & 2.9$\pm$0.2 & 21.0 & 4.1 & 11.2 & 42.5 & -1.2$^{7}$ &  \\  
{\bf 4 }  & 17:20:43.56 & -35:47:51.4 & 2.2$\times$1.1 (0.02 pc $\times$ 0.01 pc) & 36 & 3.0$\pm$0.2 & 21.4 & 4.7 & 12.8 & 22.6 & [-1.4,-1.2] & F-2 \\  
{\bf 5 }  & 17:20:43.16 & -35:47:46.6 & 1.2$\times$0.8 (0.01 pc $\times$ 0.01 pc) & -8 & 2.5$\pm$0.2 & 18.2 & 3.2 & 8.9 & 54.3 & [-2.6,-2.4],[-2.0,-1.8] & F-1 \\  
{\bf 6 }  & 17:20:44.66 & -35:46:54.3 & 1.7$\times$1.3 (0.01 pc $\times$ 0.01 pc) & 6 & 2.4$\pm$0.2 & 17.0 & 3.5 & 9.6 & 19.1 & [-3.0,-2.8],[-1.8],[-1.4,-1.2] & F-2 \\  
{\bf 7 }  & 17:20:44.51 & -35:47:0.8 & 2.0$\times$1.1 (0.02 pc $\times$ 0.01 pc) & -32 & 2.4$\pm$0.2 & 17.3 & 3.7 & 10.1 & 19.1 & -3.8,-2.8$^{7}$ & F-1(marginal)/F-3(marginal) \\  
{\bf 8 }  & 17:20:40.42 & -35:48:51.1 & 1.9$\times$1.4 (0.02 pc $\times$ 0.01 pc) & -22 & 3.0$\pm$0.2 & 21.5 & 4.7 & 12.9 & 18.5 & [-2.6,-1.4] &  \\  
{\bf 9 } & 17:20:44.48 & -35:47:17.1 & 1.9$\times$1.5 (0.02 pc $\times$ 0.01 pc) & 40 & 2.7$\pm$0.2 & 19.4 & 4.3 & 11.9 & 14.7 & [-2.8,-2.6],[-1.8,-1.2] & F-2 \\  
{\bf 10 }  & 17:20:39.27 & -35:49:16.9 & 2.1$\times$0.9 (0.02 pc $\times$ 0.01 pc) & 35 & 3.3$\pm$0.2 & 23.8 & 5.0 & 13.7 & 35.1 & [-3.8,-3.0] & F-1 \\  
{\bf 11 }  & 17:20:41.25 & -35:48:18.3 & 1.5$\times$1.0 (0.01 pc $\times$ 0.01 pc) & -39 & 2.3$\pm$0.2 & 16.6 & 3.2 & 8.7 & 28.4 & [-3.6,-2.6] &  \\  
{\bf 12 } & 17:20:39.44 & -35:49:2.8 & 2.4$\times$1.4 (0.02 pc $\times$ 0.01 pc) & -42 & 3.5$\pm$0.2 & 24.8 & 6.0 & 16.6 & 15.8 & [-3.8,-2.4] &  \\  
13  & 17:20:39.05 & -35:49:20.7 & 2.4$\times$0.9 (0.02 pc $\times$ 0.01 pc) & 19 & 1.8$\pm$0.2 & 13.1 & 2.9 & 8.1 & 15.3 & -3.8$^{7}$ &  \\  
{\bf 14 }  & 17:20:45.52 & -35:46:33.6 & 1.7$\times$1.5 (0.01 pc $\times$ 0.01 pc) & -28 & 1.7$\pm$0.2 & 12.1 & 2.6 & 7.1 & 10.7 & [-3.2,-2.8],[-2.2, -2.0], -1.2 & F-2 \\  
15$^{2}$  & 17:20:41.58 & -35:48:37.5 & 1.5$\times$1.0 (0.01 pc $\times$ 0.01 pc) & -23 & 2.6$\pm$0.2 & 18.3 & 2.1 & 7.0 & 24.4 &  & No N$_2$H$^{+}$/free-free contamination \\  
16  & 17:20:43.38 & -35:48:47.0 & 2.3$\times$0.9 (0.02 pc $\times$ 0.01 pc) & -32 & 4.5$\pm$0.2 & 32.3 & 7.1 & 19.6 & 37.8 &  & No N$_2$H$^{+}$/Out of main filament \\  
{\bf 17 } & 17:20:39.39 & -35:49:13.7 & 2.1$\times$1.7 (0.02 pc $\times$ 0.01 pc) & 6 & 3.9$\pm$0.2 & 28.1 & 6.6 & 18.2 & 17.7 & [-3.8,-2.8] &  \\  
{\bf 18 }  & 17:20:42.35 & -35:48:34.2 & 2.0$\times$1.7 (0.02 pc $\times$ 0.01 pc) & -39 & 4.8$\pm$0.2 & 34.5 & 8.1 & 22.3 & 21.9 & -3.4,[-1.6, -1.4] & free-free contamination \\  
19$^{2}$  & 17:20:40.97 & -35:48:21.2 & 1.5$\times$1.0 (0.01 pc $\times$ 0.01 pc) & -19 & 2.0$\pm$0.2 & 13.9 & 1.8 & 5.4 & 18.6 &  & No N$_2$H$^{+}$ \\  
{\bf 20 }  & 17:20:40.66 & -35:48:39.4 & 1.3$\times$1.0 (0.01 pc $\times$ 0.01 pc) & 44 & 2.1$\pm$0.2 & 15.0 & 2.8 & 7.6 & 31.0 & [-3.4,-2.6] & F-1 \\  
{\bf 21}$^{2,3}$  & 17:20:41.20 & -35:48:28.5 & 1.4$\times$0.9 (0.01 pc $\times$ 0.01 pc) & -33 & 1.1$\pm$0.1 & 8.1 & 1.4 & 3.1 & 13.4 & [-2.4,-1.2] & F-1 \\  
{\bf 22 }  & 17:20:44.39 & -35:47:26.4 & 2.1$\times$1.6 (0.02 pc $\times$ 0.01 pc) & -2 & 2.2$\pm$0.2 & 16.0 & 3.7 & 10.2 & 10.7 & [-3.8,-3.2], [-1.8,-1.2] & F-2 \\  
{\bf 23 } & 17:20:41.65 & -35:48:3.1 & 3.0$\times$1.3 (0.02 pc $\times$ 0.01 pc) & 44 & 1.7$\pm$0.2 & 12.2 & 3.2 & 8.9 & 7.0 & [-3.4,-2.4] & F-1 \\  
{\bf 24 } & 17:20:47.61 & -35:47:53.7 & 1.6$\times$1.5 (0.01 pc $\times$ 0.01 pc) & 31 & 1.2$\pm$0.2 & 8.3 & 1.7 & 4.8 & 8.1 & [-3.4,-3.2],-2.0, -1.2 & Out of main filament \\  
25  & 17:20:42.87 & -35:49:36.0 & 1.9$\times$0.8 (0.02 pc $\times$ 0.01 pc) & 34 & 1.9$\pm$0.2 & 13.7 & 2.8 & 7.6 & 23.9 &  & No N$_2$H$^{+}$/Out of main filament \\  
{\bf 26 } & 17:20:42.87 & -35:48:6.6 & 2.0$\times$0.2 (0.02 pc $\times$ 0.00 pc) & 28 & 1.5$\pm$0.2 & 10.5 & 2.0 & 5.6 & 141.9 & [-1.4,-1.2] & F-2 \\  
{\bf 27 } & 17:20:38.94 & -35:49:5.9 & 2.1$\times$1.0 (0.02 pc $\times$ 0.01 pc) & -19 & 1.9$\pm$0.2 & 13.6 & 2.9 & 7.9 & 17.0 & [-2.2,-1.6] &  \\  
28  & 17:20:41.88 & -35:49:1.6 & 3.4$\times$0.9 (0.03 pc $\times$ 0.01 pc) & -41 & 2.2$\pm$0.2 & 16.0 & 4.3 & 11.9 & 14.4 &  & No N$_2$H$^{+}$/Out of main filament \\  
29$^{2,3}$  & 17:20:39.59 & -35:48:42.6 & 1.4$\times$0.9 (0.01 pc $\times$ 0.01 pc) & 15 & 0.6$\pm$0.1 & 4.5 & 0.3 & 1.7 & 7.4 &  & No N$_2$H$^{+}$ \\  
30 & 17:20:42.35 & -35:49:31.8 & 1.7$\times$1.7 (0.01 pc $\times$ 0.01 pc) & 1 & 2.7$\pm$0.2 & 19.1 & 4.2 & 11.6 & 14.6 &  & No N$_2$H$^{+}$/Out of main filament \\  
{\bf 31 }  & 17:20:44.42 & -35:47:21.2 & 2.2$\times$1.5 (0.02 pc $\times$ 0.01 pc) & 22 & 2.1$\pm$0.2 & 15.2 & 3.5 & 9.7 & 10.7 & -3.8, [-2.0, -1.8] & F-2 \\  
32$^{2,3}$  & 17:20:44.26 & -35:49:14.1 & 1.4$\times$0.9 (0.01 pc $\times$ 0.01 pc) & 39 & 3.3$\pm$0.1 & 23.4 & 1.1 & 9.0 & 38.5 &  & No N$_2$H$^{+}$/Out of main filament \\  
33$^{2}$  & 17:20:43.39 & -35:47:57.3 & 1.5$\times$1.0 (0.01 pc $\times$ 0.01 pc) & 42 & 2.3$\pm$0.2 & 16.4 & 0.7 & 6.3 & 21.9 & -1.8$^{7}$ & F-2(marginal) \\  
{\bf 34 } & 17:20:39.27 & -35:47:59.3 & 2.3$\times$1.3 (0.02 pc $\times$ 0.01 pc) & 34 & 2.0$\pm$0.2 & 14.4 & 3.4 & 9.2 & 10.7 & [-3.8, -3.6, -3.4], [-2.8, -2.6] & Out of main filament  \\  
35$^{2,3}$  & 17:20:38.34 & -35:48:32.4 & 1.4$\times$0.9 (0.01 pc $\times$ 0.01 pc) & 15 & 0.9$\pm$0.1 & 6.7 & 1.2 & 2.6 & 11.0 & -2.2$^{7}$ & Out of main filament  \\  
36  & 17:20:42.92 & -35:49:24.7 & 2.1$\times$1.9 (0.02 pc $\times$ 0.02 pc) & -19 & 3.8$\pm$0.2 & 27.0 & 6.7 & 18.5 & 14.1 &  & No N$_2$H$^{+}$/Out of main filament \\  
37  & 17:20:36.73 & -35:48:56.6 & 1.9$\times$1.4 (0.02 pc $\times$ 0.01 pc) & 38 & 1.4$\pm$0.2 & 9.9 & 2.2 & 6.0 & 8.4 &  & No N$_2$H$^{+}$/Out of main filament \\  
38  & 17:20:43.44 & -35:49:6.4 & 2.2$\times$1.8 (0.02 pc $\times$ 0.01 pc) & -4 & 1.8$\pm$0.2 & 12.9 & 3.1 & 8.7 & 7.2 &  & No N$_2$H$^{+}$/Out of main filament \\  
{\bf 39 } & 17:20:40.25 & -35:48:46.7 & 2.5$\times$1.5 (0.02 pc $\times$ 0.01 pc) & -36 & 2.3$\pm$0.2 & 16.4 & 4.0 & 11.1 & 10.0 & [-2.6,-2.2] &  \\  
40  & 17:20:42.82 & -35:49:20.4 & 2.0$\times$1.9 (0.02 pc $\times$ 0.02 pc) & 0 & 3.4$\pm$0.2 & 24.4 & 5.9 & 16.3 & 13.8 &  & No N$_2$H$^{+}$/Out of main filament  \\  
\hline 
\end{tabular} 
} 
\tablefoot{
\tablefoottext{$^{1}$}{ID numbers of 3.1 mm compact sources associated with N$_2$H$^+$ emission in two or more contiguous channels are labeled in bold.}\\ 
\tablefoottext{$^{2}$}{Deconvolved source size are not obtained since the structure is not resolved. The peak flux was used to estimate the mass.}\\ 
\tablefoottext{$^{3}$}{Results of the 2D Gaussian fitting on the 12m+7m map. For IDs 21, 29, 32, 35, the 12m map was used since the 3.1mm continuum emission is not detected with a $>$3$\sigma$ on the 12m+7m map. }\\ 
\tablefoottext{$^{4}$}{Deconvolved FWHM source diameters along the major and minor axes.}\\  
\tablefoottext{$^{5}$}{We adopted $T_{\rm dust}$ = 20 K for all sources, except for source ID 2 for which we assumed $T_{\rm dust}$ = 50 K (see Sect. \ref{sect:getsources}).}  \\
\tablefoottext{$^{6}$}{Volume-averaged gas density derived assuming spheroidal cores as in Eq.~(2).}\\
\tablefoottext{$^{7}$}{Marginal N$_2$H$^+$ detection since it is detected in only one channel for this source.}
} 
\end{table*}

\begin{figure*}
\centering
\includegraphics[width=180mm, angle=0]{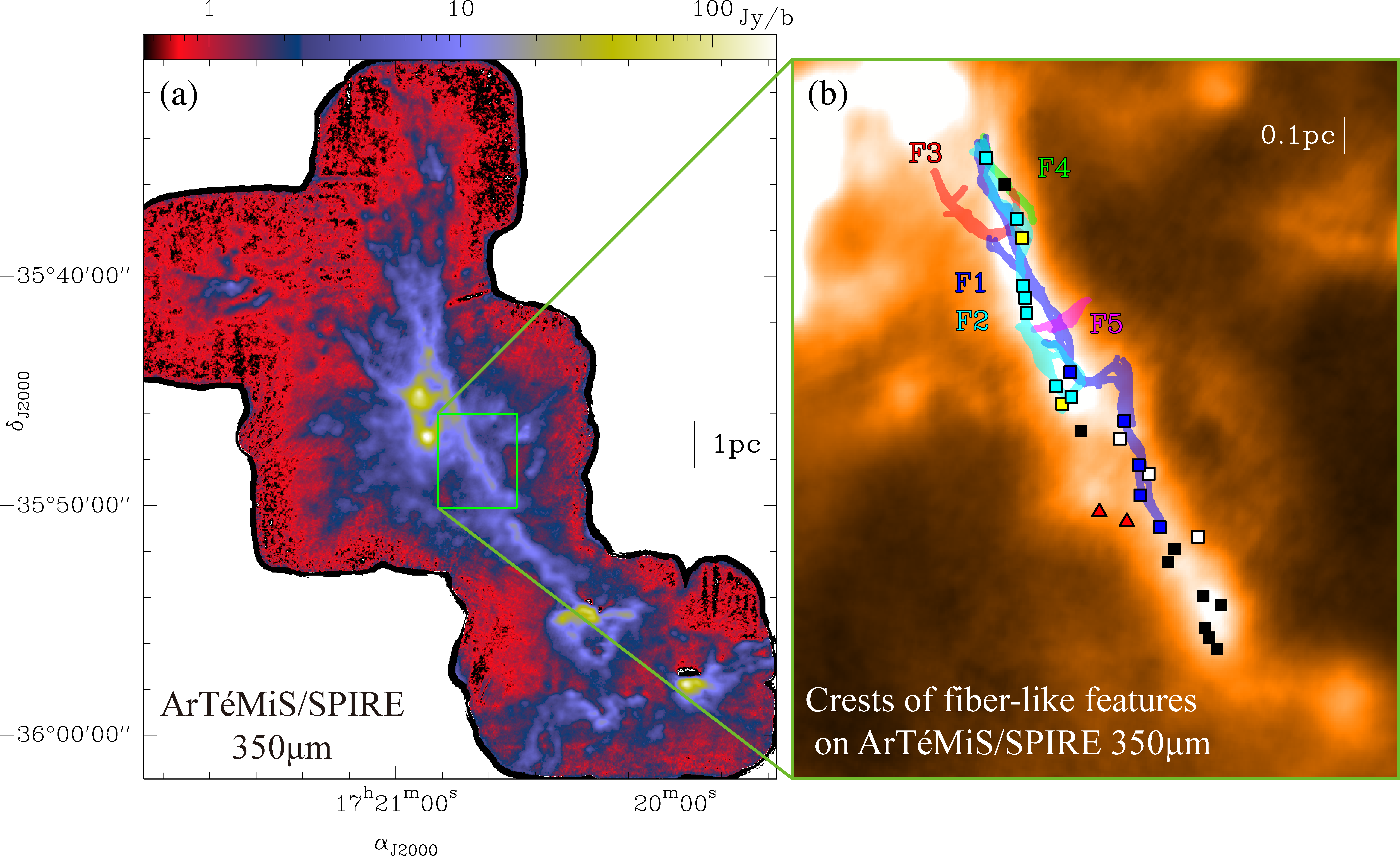}
\caption{({\it a}) ArT$\acute{\rm e}$MiS 350 $\mu$m dust continuum map of the NGC~6334 region \citep{Andre16} 
and ({\it b}) blow-up of the field imaged here with ALMA, with the skeleton of the velocity-coherent fiber-like structures 
identified in the N$_2$H$^+$ data cube overlaid as colored curves 
(see Sect. \ref{sect:Post-selection} and Table \ref{table:N$_2$H$^+$_filament}). 
In panel ({\it b}), 
dark blue squares mark 3.1 mm compact sources associated with N$_2$H$^+$ emission and embedded in F-1, 
light blue squares 3.1 mm sources associated with N$_2$H$^+$ emission and embedded in F-2,
yellow squares 3.1 mm sources embedded in F-1, F-2, or F-3 and marginally detected in only one N$_2$H$^+$ channel,
black squares 3.1 mm sources associated with N$_2$H$^+$ emission in the main filament but not embedded in any N$_2$H$^+$ velocity-coherent structure,
and white squares 3.1 mm sources in the main filament not associated with any N$_2$H$^+$ emission.
The red triangles indicate 3.1 mm compact sources possibly contaminated by free-free emission. 
}
\label{figs:getfilament}
\end{figure*}

\begin{figure*}
\centering
\includegraphics[width=160mm, angle=0]{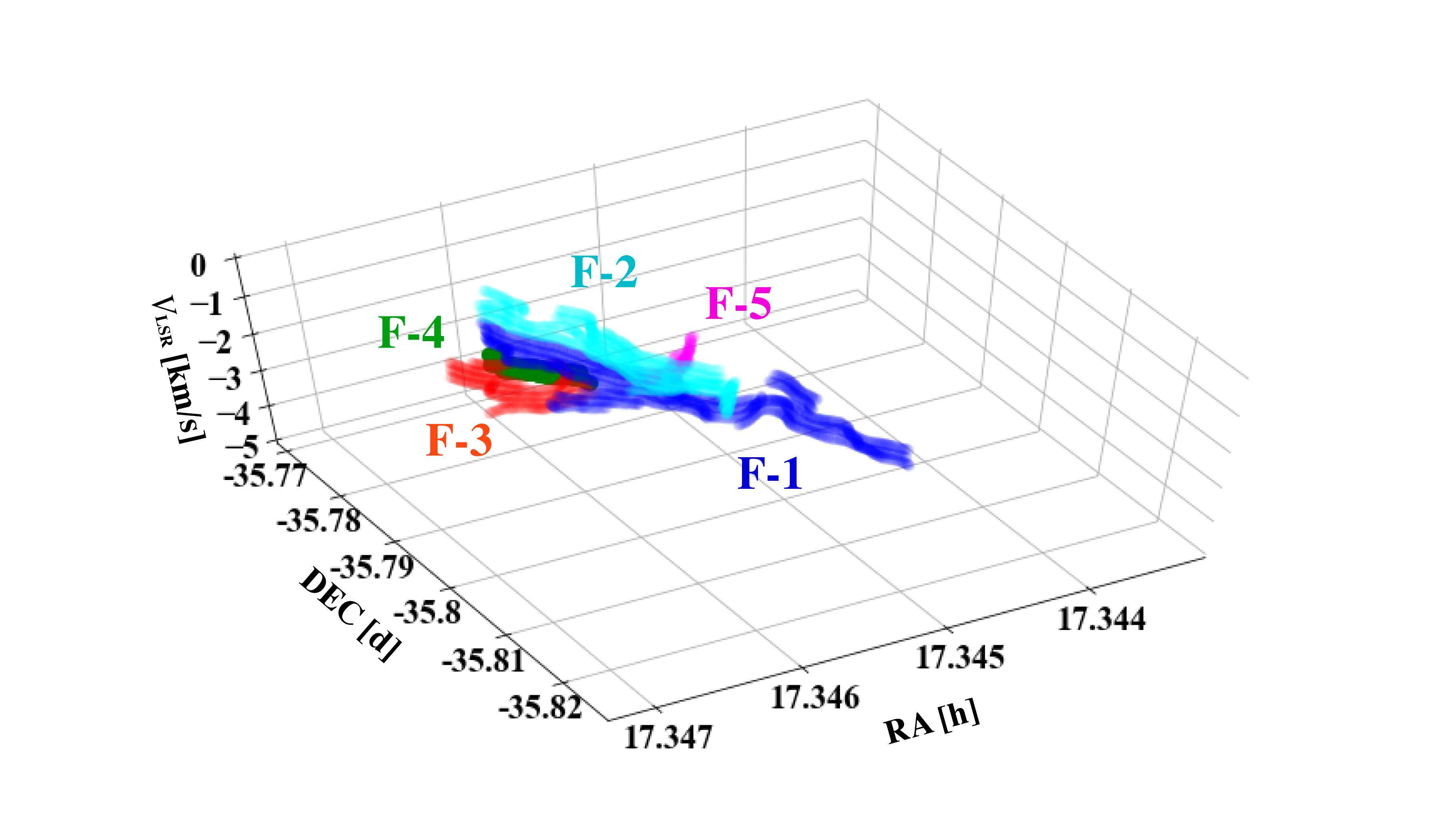}
\caption{N$_2$H$^+$(1--0) Position-Position-Velocity plot of the NGC~6334 filament system. 
Colored curves indicate the crests of the N$_2$H$^+$ velocity-coherent structures identified with getfilaments, after applying the 
post-extraction selection criteria described in Sect.\ref{sect:Post-selection}, and using the same color coding as in Fig.~\ref{figs:getfilament}b.}
\label{figs:3d_skelton}
\end{figure*}

\subsubsection{Compact source extraction from the ALMA 3.1 mm continuum data with \textsl{getsources}}\label{sect:getsources}

To identify compact sources in the ALMA 3.1 mm continuum map, we applied the \textsl{getsources} algorithm \citep[e.g.,][]{Menshchikov12}. 
\textsl{getsources} is a multi-scale source extraction algorithm primarily developed for the exploitation of multi-wavelength far-infrared and 
submillimeter continuum data resulting from {\it Herschel} surveys of Galactic star-forming regions \citep[see ][]{Konyves15}, 
but can also be used with single-band continuum data and spectral line data.  
Source extraction with this algorithm has only one free parameter, namely the maximum size of the structures to be extracted from the images. 
Here, we adopted a maximum size of 15 $\arcsec$ (or $\sim$0.12 pc), which is comparable to the transverse full width at half maximum (FWHM) 
size of the filament as measured with ArT\'eMiS \citep{Andre16}.

After running \textsl{getsources} on the 12m-only data\footnote{We used the 12m-only instead of the 12m$+$7m image 
as the sensitivity of the data is better for the detection of compact ($< 0.1\,$pc) sources (cf. Table~\ref{table:cont_getsources_12m_only}).} 
and applying post-extraction selection criteria (see Appendix \ref{Sect:getsources_post_selection_detail}), 
we identified a total of 40 candidate compact 3.1 mm continuum sources. 
As shown in Fig. \ref{figs}$a$, 28 of these 40 sources are located within the main filament 
defined as the intersection of the area within $\pm$30$\arcsec$ (or $\pm$0.25\,pc) 
from the filament crest 
and the area enclosed within the 5 Jy beam$^{-1}$ contour in the ArT$\acute{\rm e}$MiS 350 $\mu$m map 
(see also Fig. \ref{figs:ArTeMiS_ALMA_core_association} and Table~\ref{table:cont_getsources_12m_only}). 
Two of these 28 sources (IDs 15, 18 in Table \ref{table:cont_getsources_12m_only}) are probably affected by contamination 
from free-free emission as mentioned in Sect.~\ref{result:cont}. 
The positions and sizes of all of these 40 compact continuum sources are summarized in Table~\ref{table:cont_getsources_12m_only}, 
along with their basic properties.

To investigate whether the compact continuum sources identified above are associated with N$_2$H$^+$ emission, 
we also applied \textsl{getsources} to each N$_2$H$^+$ velocity channel map (see Appendix~\ref{Sect:getsources_post_selection_detail} for details). 
We found that 23 of the 40 compact continuum sources are associated with N$_2$H$^+$ emission in at least two consecutive channels. 
In addition, five of the 40 compact sources (IDs 3, 7, 13, 33, 35 in Table \ref{table:cont_getsources_12m_only}) are associated 
with N$_2$H$^+$ (1--0) line emission in only one channel, but these sources did not pass our post-extraction selection criteria 
(see Appendix \ref{Sect:getsources_post_selection_detail}).  
Although contamination of the 3.1 mm continuum data by free-free emission is an issue, compact sources robustly detected in both 3.1 mm continuum 
and N$_2$H$^+$ emission are unlikely to be affected. 

We estimated the mass ($M_{\rm tot}$) of each compact continuum source, 
under the assumption that all of the 3.1 mm continuum emission arises from dust and that the emission is optically thin. 
The mass 
was obtained from the integrated 3.1 mm flux density $S_{\rm 3.1mm}^{\rm tot}$ 
derived from two-dimensional Gaussian fitting in the 12m$+$7m continuum image\footnote{We used the 12m$+$7m data 
to estimate the integrated flux densities and masses of the sources detected in the 12m-only continuum map, in order to avoid missing-flux problems 
due to interferometric filtering of large scales as much as possible. For four weak unresolved sources, undetected above the $3\sigma $ level 
in the 12m$+$7m map, the 12m-only data were used instead (cf. Table~\ref{table:cont_getsources_12m_only}).} 
using the formula: 

\begin{equation}\label{eq1}
M_{\rm tot} = \frac{S_{\rm 3.1mm}^{\rm tot} d^2}{\kappa_{\rm 3.1mm} B_{\rm 3.1mm}(T_{\rm d})}, 
\end{equation}

\noindent where $d$, $\kappa_{\rm 3.1mm}$, and $B_{\rm 3.1mm}$($T_{\rm d}$) are the distance to the target, the dust opacity (per unit mass of gas + dust), 
and the Planck function at dust temperature $T_{\rm d}$, respectively. 
We adopted the same dust opacity law as the {\it Herschel} Gould Belt survey team, 
namely $\kappa_\lambda$ = 0.1 ($\lambda$/300$\mu$m)$^{-\beta}$ cm$^2$g$^{-1}$ \citep[cf.][]{Hildebrand83,Roy14} with $\beta =2$ and here $\lambda = 3.1\,$mm. 
If $\beta = 1.5$ instead of $\beta = 2$, the core masses would be a factor of $\sim$3 lower than the values listed in Table~\ref{table:cont_getsources_12m_only}.
For most sources, we adopted $T_{\rm dust}$ = 20 K which corresponds to the median dust temperature derived from {\it Herschel} data along the crest of the filament \citep{Andre16,Tige17}. 
For source ID 1, we adopted a higher temperature value ($T_{\rm dust}$ = 50 K) as this object coincides with a bright {\it Spitzer} 8 $\mu$m source 
and is most likely an internally-heated, relatively massive protostellar core. 
The 5$\sigma$ mass sensitivity of the ALMA 12m continuum data for compact sources 
is $\sim$$2.0\, M_{\odot}$, corresponding to $S_{\rm 3.1mm}^{\rm tot} =0.7$~mJy, assuming $\beta$ = 2 and $T_{\rm dust}$ = 20 K.
The average gas density ($\equiv \overline{n_{\rm H_2}}$) of each source was then derived assuming spheroidal cores as follows:

\begin{equation}
\overline{n_{\rm H_2}} = \frac{M_{\rm tot}}{\frac{4}{3}\pi[\sqrt{R_{\rm major} R_{\rm minor}}]^3}, 
\end{equation}

\noindent where $R_{\rm major}$ and $R_{\rm minor}$ are the deconvolved FWHM sizes along the major and minor axis of the source, respectively \citep[cf.][]{Konyves15}. 
The median mass of the 26 compact continuum sources embedded in the main filament\footnote{The median mass of all 40 compact continuum sources 
is 9.4$^{+3.4}_{-1.9}$ $M_{\odot}$ (lower quartile: 7.5 $M_{\odot}$, upper quartile: 12.8 $M_{\odot}$) and the median
volume-averaged density 1.4$\times$10$^{7}$ cm$^{-3}$ 
(lower quartile: 0.8$\times$10$^{7}$ cm$^{-3}$, upper quartile: 2.0$\times$10$^{7}$ cm$^{-3}$).}
is $9.6^{+3.0}_{-1.9}$ $M_{\odot}$ (lower quartile: 7.7 $M_{\odot}$, upper quartile: 12.6 $M_{\odot}$)
and their median volume-averaged density 1.6$\times$10$^{7}$ cm$^{-3}$ (lower quartile: 1.0$\times$10$^{7}$ cm$^{-3}$, 
upper quartile: 2.2$\times$10$^{7}$ cm$^{-3}$). 
The 26 continuum sources embedded in the main filament are compact (with an estimated typical outer radius $\sim 5000\,$au)
and have a very high volume-averaged density ($\sim$10$^{7}$ cm$^{-3}$), 
suggesting that they are on their way to form stars.
Moreover, their spatial distribution closely follows that of the ArT$\acute{\rm e}$MiS 350$\mu$m emission clumps 
(see Fig.~\ref{figs:ArTeMiS_ALMA_core_association} and Sect.~\ref{two_levels}). 
In the following, we therefore regard these 26 compact 3.1 mm continuum sources as dense cores.

\subsubsection{Extraction of fiber-like structures from the N$_2$H$^+$ data cube with \textsl{getfilaments}}

\label{sect:subtraction_uncomp}

{\sl Removal of unrelated velocity components.} In the NGC~6334 region, we found evidence of the presence of several velocity components along the line of sight (see Sect. \ref{Sect:velo_structure}). 
To avoid contamination of the N$_2$H$^+$(1--0) HFS~1 line emission from the main velocity component at $V_{\rm sys}$=$-$2.6 km s$^{-1}$ 
by the HFS~2-7 emission from other velocity components, we fit the N$_2$H$^+$ emission observed at each pixel with a multiple-velocity-component HFS model 
and subtracted the corresponding HFS~2-7 emission from the data cube.
As shown in Fig. \ref{figs:N2Hp_contamination} in Appendix \ref{Sect:Contamination}), if the velocity width of each component is 
$\sim$1.0 km s$^{-1}$, then the N$_2$H$^+$ HFS~1 emission of the main component can be contaminated 
by the HFS~2-7 emission of a more blue-shifted velocity component when the velocity difference between the two components 
along the line of sight is $\sim \,$5.0-7.0 km s$^{-1}$.  
Figure~\ref{figs:n2h+_spectrum_removal} shows four examples of N$_2$H$^+$(1--0) spectra obtained after subtraction of contaminating components.
Hereafter, we use the subtracted data cube in our scientific analysis. 

\label{sect:Post-selection}

\begin{figure*}
\centering
\includegraphics[width=190mm, angle=0]{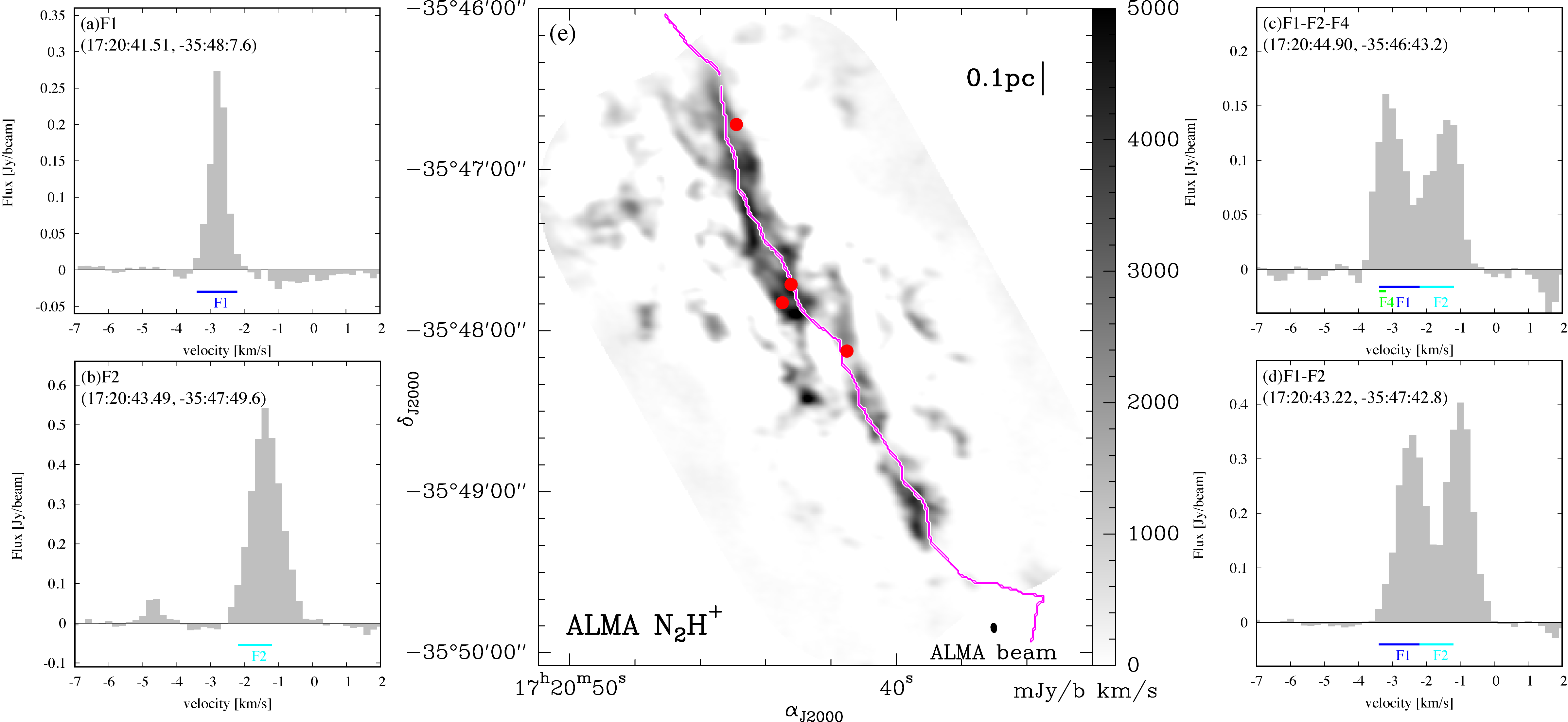}
\caption{Four examples of spectra showing the isolated component of the N$_2$H$^+$(1--0) multiplet (HFS 1)
in the original data cube (i.e. before subtracting unrelated velocity components). 
The positions of these spectra are indicated at the top-left corner of each panel. 
The central panel is the N$_2$H$^+$(1--0) intensity map (integrated from $-$11.4 km s$^{-1}$ to $+$16.8 km s$^{-1}$ -- cf. Fig.~\ref{figs}b). 
The red filled circles in the central panel also mark the positions 
of the four spectra. Colored horizontal segments indicate the velocity ranges of the various fiber-like structures (see Table \ref{table:N$_2$H$^+$_filament}).
}
\label{figs:n2h+_fiber_spectrum}
\end{figure*}

\begin{table} 
\centering 
\caption{N$_2$H$^+$ "fiber-like" structures identified with {\it getfilaments} \label{table:N$_2$H$^+$_filament}} 
\begin{tabular}{lccc} 
\hline 
ID &  length & velocity range &  \\ 
\hline 
F-1 &  1.21 pc & [$-$3.4,$-$2.2 km/s]  &   \\ 
F-2 &  0.72 pc & [$-$2.2,$-$1.2 km/s]  &   \\ 
F-3 &  0.33 pc & [$-$3.6,$-$2.4 km/s]  &   \\ 
F-4 &  0.26 pc & [$-$3.4,$-$3.2 km/s]  &   \\ 
F-5 &  0.17 pc & [$-$2.2,$-$1.6 km/s]  &   \\ 
\hline 
\end{tabular} 
\end{table}

\begin{figure*}
\centering
\includegraphics[width= 180mm, angle=0]{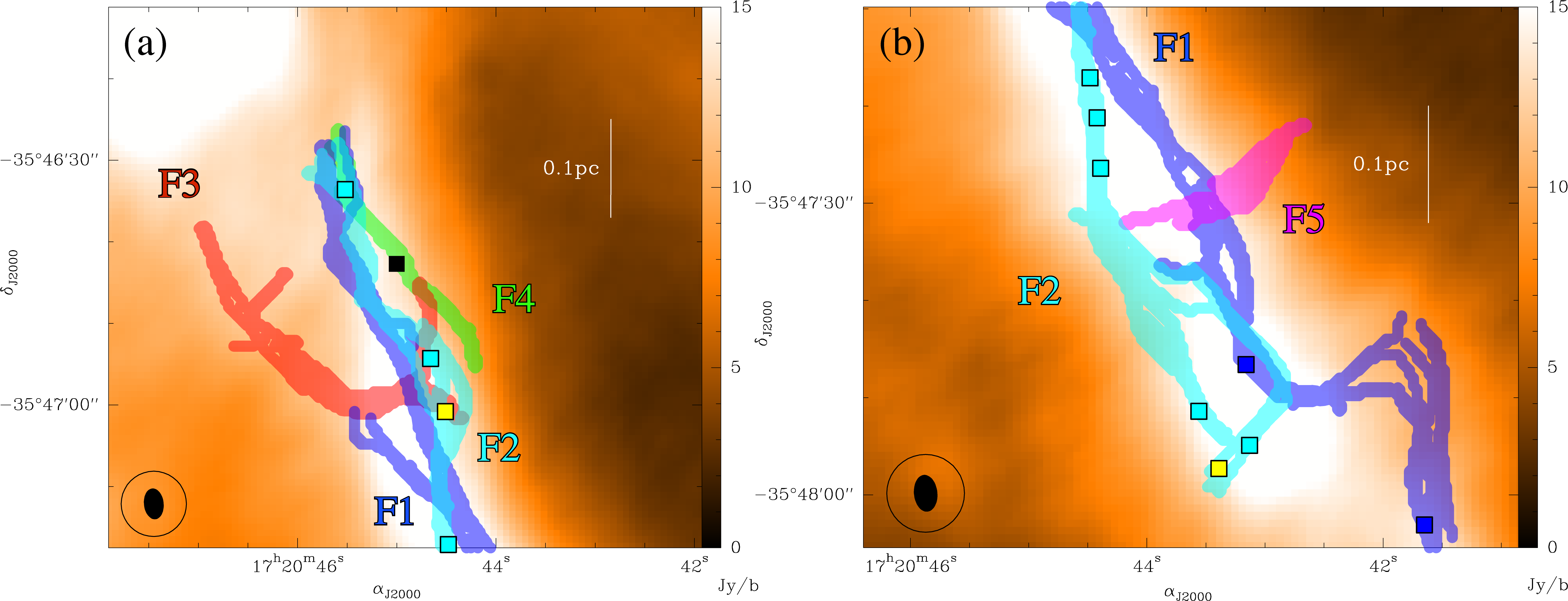}
\caption{Close-up views of the northern ({\it a}) and southern ({\it b}) parts of the dusty filament seen in the ArT$\acute{\rm e}$MiS 350 $\mu$m map. 
Symbols and curves are the same as in Fig. \ref{figs:getfilament}b. 
The ALMA 3.1 mm continuum and ArT$\acute{\rm e}$MiS 350 $\mu$m beams are shown as a filled ellipse and an open circle at the bottom left of each panel.}
\label{figs:getfilament_zoom}
\end{figure*}

{\sl Extraction of velocity-coherent fiber-like features with \textsl{getfilaments}.} We extracted filamentary structures from the ALMA N$_2$H$^+$ data cube after subtraction of compact N$_2$H$^+$ emission using the \textsl{getfilaments} algorithm \citep[e.g.,][]{Menshchikov13}. 
Each N$_2$H$^+$ velocity channel map was decomposed into a set of spatially-filtered single-scale images from small (3$\arcsec$) to large (16$\arcsec$) scales. 
At each scale, the algorithm separated filamentary structures and compact sources, 
and extracted significant filamentary structures in the source-subtracted image component.
All extracted filamentary structures were merged into a single image per velocity channel map. 
We selected only filamentary structures whose $>$90\% pixels along the crest were detected above the $>$ 3$\sigma$ level in each channel.
As the velocity resolution of our ALMA N$_2$H$^+$ data (0.2 km s$^{-1}$, that is, less than the isothermal sound speed in the cloud) 
should be sufficient to resolve the velocity width of the filament, 
we also imposed that a selected filamentary structure should be detected in at least two consecutive velocity channels. 
By comparing velocity-channel maps, we therefore associated filamentary structures with matching spatial distributions among 
channels (see Appendix \ref{Sect:getfilament_post_selection_detail}). 
Our procedure is similar to that used by \citet{Hacar13} to trace velocity-coherent structures in data cubes, in the sense that 
both methods work in position-position-velocity (PPV) space\footnote{As only a small ($\geq 5$) number 
of velocity-coherent features appeared to be present in our N$_2$H$^+$ data cube (see Table~\ref{table:N$_2$H$^+$_filament}) 
and we had no experience with the FIVE method developed by \citet{Hacar13}, we preferred to follow a procedure based 
on channel-by-channel \textsl{getfilaments} extractions and careful visual inspection of the data cube (see Appendix \ref{Sect:getfilament_post_selection_detail}).}.
In this way, we identified five distinct fiber-like structures, labeled F-1 to F-5, whose crests are displayed in Fig.~\ref{figs:getfilament} and Fig.~\ref{figs:3d_skelton}. 
Examples of individual N$_2$H$^+$ (HFS 1) spectra across the two main fiber-like structures F-1 and F-2 are shown in Fig. \ref{figs:n2h+_fiber_spectrum}. 
The spatial distribution of the structure labeled F-2 partly overlaps with the distribution of F-1 at $-$2.4 km s$^{-1}$. 
But the northern portion of F-2 lies slightly to the west of F-1, while the southern part of F-2 lies to the east of F-1 (see, e.g., Fig.~\ref{figs:getfilament}b). 
Moreover, at positions where F-1 and F-2 overlap in the plane of the sky, the N$_2$H$^+$ spectra clearly exhibit distinct velocity components  [see, e.g., positions (c) 
and (d) in Fig.~\ref{figs:n2h+_fiber_spectrum}]. Therefore, F-1 and F-2 seem to be distinct velocity-coherent features.
The five 
fiber-like structures are likely associated with the dust continuum filament seen in the ArT$\acute{\rm e}$MiS 350 $\mu$m map.
Similar sub-filamentary structures have been found in the low-mass star-forming filament B211/B213 in Taurus \citep{Hacar13} 
and have been called fibers in the literature \citep[cf.][]{Tafalla15}. 
At this point, we refer to the five features F-1--F-5 as 
fiber-like structures.
(In Sect~\ref{fibers} below, we argue that only two of them, F-1 and F-2, may be genuine fibers, that is velocity-coherent sub-structures 
of the main filament itself.)
Their typical length 
is 0.5$\pm$0.4 pc (see Table \ref{table:N$_2$H$^+$_filament}, which also gives their velocity ranges). 

A portion of the dusty filament seen in the ArT$\acute{\rm e}$MiS 350 $\mu$m continuum map 
was not traced by \textsl{getfilaments} in the N$_{2}$H$^{+}$ data, especially in the southern part of the field in Fig.~\ref{figs}
although the filament can be recognized by eye in the N$_2$H$^+$(1--0)  integrated intensity map (Fig.\ref{figs}$b$). 
The reason why this southern portion was not traced in the N$_2$H$^+$ velocity channel maps with \textsl{getfilaments}  
is that the N$_2$H$^+$ emission becomes very clumpy in this area 
and was subtracted out as a collection of point-like sources 
by \textsl{getsources} before identification of filamentary structures. 

F-1 is the longest and F-2 the second longest of the five extracted velocity features. 
Both of these fiber-like structures are roughly parallel to the dust continuum filament seen in the ArT$\acute{\rm e}$MiS 350 $\mu$m map. 
In contrast, the crest of F-5 is roughly perpendicular to the dust filament. 
In the northern part of the N$_2$H$^+$ map (north of -35$^{\circ}$47$\arcmin$10$\arcsec$ in Fig. \ref{figs:getfilament} and Fig. \ref{figs:getfilament_zoom}a), 
the spatial distributions of F-1, F-2, F-4  partly overlap in the plane of sky. 
The three features F-1, F-2, F-4  are nevertheless separated from each other in velocity space (see Table~\ref{table:N$_2$H$^+$_filament}).
The crests of F-1, F-2, F-4 are also slightly shifted from one another by typically one ALMA beam  ($\sim$3$\arcsec$, see Fig. \ref{figs:getfilament_zoom}a). 
In the northern part, they are distributed in the sequence F-1, F-2, F-4 from east to west (see Fig. \ref{figs:getfilament_zoom}a). 
In contrast, the projected LSR velocities are in the order F-4 ($\sim$ $-$3.3 km s$^{-1}$), F-1 ($\sim$ $-$2.8 km s$^{-1}$), F-2 ($\sim$ $-$1.7 km s$^{-1}$). 
Assuming F-1, F-2, F-4 are part of the same filament, 
this different ordering between the spatial and velocity distributions cannot be explained
by a simple transverse velocity gradient across the parent filament due to either accretion onto and/or rotation of the filament.
In the southern part of the field (south of -35$^{\circ}$47$\arcmin$10$\arcsec$ in Fig. \ref{figs:getfilament} and Fig. \ref{figs:getfilament_zoom}b), F-4 is not detected
and F-1, F-2 are distributed in the order F-2, F-1 from east to west (see Fig. \ref{figs:getfilament_zoom}b).  
These opposite spatial configurations for F-1, F-2 between the northern and the southern part of the dust filament 
are suggestive of an intertwined, double helix-like pattern (DNA-like) (see Fig. \ref{figs:DNA_fiber}b).

We also extracted filamentary structures from the ALMA HC$_3$N, HC$_5$N, CH$_3$CCH, and H$_2$CS integrated intensity maps using \textsl{getfilaments}. We detected one filamentary structure in HC$_3$N, one filamentary structure in HC$_5$N, one filamentary structure in CH$_3$CCH, and two filamentary structures in H$_2$CS. 
These filamentary structures roughly correspond to the F-1  and/or F-2  "fiber-like" features identified in N$_2$H$^+$. Due to the lower effective sensitivity of the H$_2$CS(3$_{1,2}$-2$_{1,1}$) data, the NGC 6334 filament is broken up into two segments in H$_2$CS. The H$_2$CS filamentary structures coincide with the northern and southern parts of the filamentary structures traced in other lines. Owing to the lower line intensities and/or lower spectral resolution of our data in HC$_3$N, HC$_5$N, CH$_3$CCH, and H$_2$CS, we could not carry out as detailed a PPV analysis in these lines as in N$_2$H$^+$.


\begin{figure*}
\centering
\includegraphics[width=180mm, angle=0]{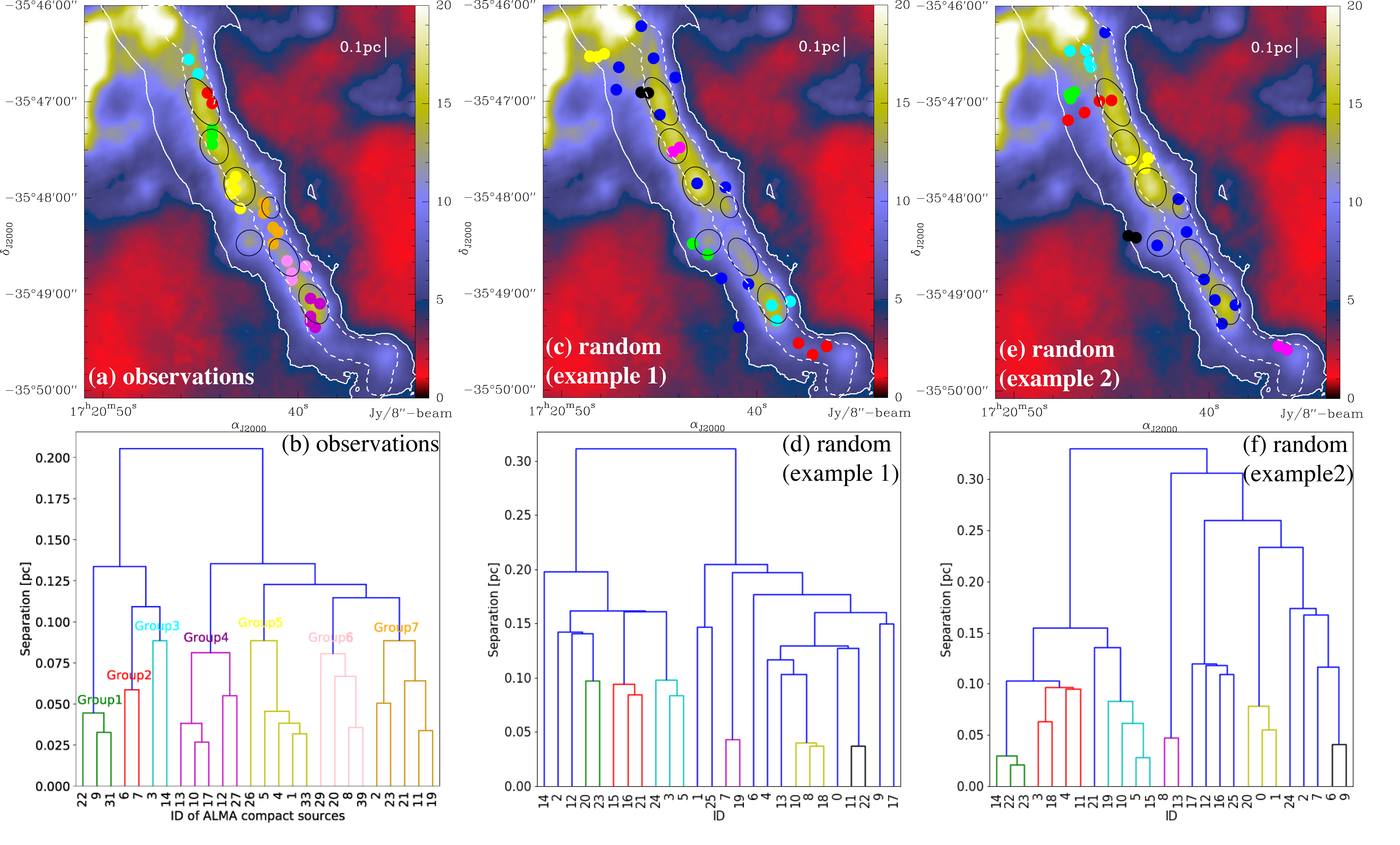}
\caption{Comparison of the spatial distributions and NNS analysis results 
obtained for the observed ALMA cores (left column) with those obtained for two examples of randomly-distributed source populations (middle and right columns of panels). The top row shows the distributions of observed cores (panel a) and randomly-placed sources (panels c and e) overlaid on the ArTeMiS 350$\mu$m continuum map (color scale). 
In each panel, the solid white contour marks the footprint of the main filament, defined as the intersection of the area within $\pm$30$\arcsec$ from the filament crest and the interior of the 5~Jy\,beam$^{-1}$ contour in the ArTeMiS 350 $\mu$m map, within which randomly-placed sources were distributed. 
The  dashed white contour marks the area within $\pm$6$\arcsec$ (or 0.05 pc) of the filament crest, defining the inner filament footprint also discussed in the text. The bottom row shows the NNS analysis results for the observed cores (panel b) and the randomly-placed sources (panels d and f). Colored circles mark core positions in panels (a), (c), (e), with colors corresponding to the groups defined by the NNS dendrograms shown in panels (b), (d), (f).
}
\label{figs_NNStest}
\label{figs:NNS_result}
\end{figure*}

\begin{table*} 
\tiny
\centering  
\caption{Properties of ArT$\acute{\rm e}$MiS dense clumps \label{table:ArTeMiS_clumps}}  
\scalebox{0.9}{ 
\begin{tabular}{lcccccccccc} 
\hline 
ArT$\acute{\rm e}$MiS &  RA & DEC & $R_{\rm major} \times R_{\rm minor}$$^\dag$  &PA & $S_{350\mu m}^{\rm peak}$ & $S_{350\mu m}^{\rm tot}$ & $M_{\rm clump}$$^\ddag$  & $\overline{n_{\rm H_2}}$ & $\sigma_{\rm N_2H^+}$ & IDs of  \\   
clump  & [h:m:s] & [d:m:s] &   & [deg] & [Jy/beam] & [Jy] & [$M_{\odot}$] & [$\times$10$^5$ cm$^{-3}$]  & [km/s]& associated ALMA cores \\
\hline 
1 & 17:20:44.9 & -35:46:60 & 31.$\arcsec$9{$\times$}13.$\arcsec$7 (0.26pc {$\times$}0.11pc) & 15 & 17 & 227 & 320 &2.1 & 0.5 & \textcolor{black}{6,7 (Group2)} \\
2 & 17:20:44.3 & -35:47:28 & 21.$\arcsec$2{$\times$}14.$\arcsec$8 (0.17pc {$\times$}0.12pc) & 22 & 11 & 106 & 149 &1.7 & 0.8 & \textcolor{black}{9,19,24 (Group1)} \\
3 & 17:20:43.0 & -35:47:53 & 23.$\arcsec$6{$\times$}16.$\arcsec$9 (0.19pc {$\times$}0.14pc) & 24 & 18 & 216 & 303 &2.3 & 1.0 & \textcolor{black}{1,4,5,21,25 (Group5)} \\
4 & 17:20:41.4 & -35:48:06 & 10.$\arcsec$9{$\times$}7.$\arcsec$5 (0.09pc {$\times$}0.06pc) & 26 & 9 & 36 & 51 &4.2 & 0.3 & \textcolor{black}{2,11,16,18,20 (Group7)} \\
5 & 17:20:42.5 & -35:48:28 & 14.$\arcsec$7{$\times$}13.$\arcsec$7 (0.12pc {$\times$}0.11pc) & -80 & 12 & 82 & 116 &2.5 & 0.8 & \\
6 & 17:20:40.7 & -35:48:37 & 24.$\arcsec$6{$\times$}13.$\arcsec$0 (0.20pc {$\times$}0.11pc) & 13 & 11 & 117 & 164 &1.8 & 0.4 & \textcolor{black}{8,17,23,26 (Group6)} \\
7 & 17:20:39.2 & -35:49:06 & 25.$\arcsec$3{$\times$}14.$\arcsec$0 (0.21pc {$\times$}0.12pc) & 19 & 16 & 177 & 249 &2.3 & 0.5 & \textcolor{black}{10,12,13,15,22 (Group4)} \\
\hline 
\end{tabular} 
}
\tablefoot{\tablefoottext{\dag}{Deconvolved \textcolor{black}{FWHM} source diameters along the major and minor axes. }\\
\tablefoottext{\ddag}{\textcolor{black}{Estimated using an equation similar to Eq. (\ref{eq1}) at $\lambda$ = 350 $\mu$m assuming $T_{\rm dust}$ = 20 K.}}}
\end{table*}


\section{Discussion}\label{Sect:Discussion}

Our ALMA study has revealed, for the first time, the presence of density and velocity sub-structure in the NGC~6334 filament. 
We discuss the two types of sub-structure in turn in the following. 

\subsection{Bimodal fragmentation in the NGC 6334 filament}\label{two_levels}

While it was not clear whether dense cores were embedded in the NGC 6334 filament based only on the ArT$\acute{\rm e}$MiS 350 $\mu$m data \citep{Andre16}, 
our ALMA 3.1 mm continuum map with an angular resolution of $\sim$$2.3\arcsec$ has allowed us to identify 26 candidate dense cores within the filament (cf. Sect. \ref{sect:getsources}). These ALMA cores appear to be clustered in several groups closely associated with ArT$\acute{\rm e}$MiS clumps (see Fig. \ref{figs:ArTeMiS_ALMA_core_association}).
Figure~\ref{figs:NNS_result} shows a dendrogram tree obtained when applying a nearest neighbor separation (NNS) analysis
to the population of ALMA dense cores (with the {\it scipy} function {\it cluster.hierarchy.linkage} and the {\it single} method). 
Here, we adopted two values of the NNS grouping threshold, 0.1~pc and 0.15~pc, 
consistent with both the typical inner width of the filament and the typical FWHM size of the ArT$\acute{\rm e}$MiS clumps (see Table \ref{table:ArTeMiS_clumps}).
Based on the NNS analysis with a grouping threshold of 0.1 pc, the 26 dense cores can be divided into seven groups of cores (groups 1--7). 
Using a grouping threshold of 0.15 pc instead of 0.1 pc,  
groups 2, 3 
join group 1 and groups 4, 5, 6, 7 
merge into the same group (see Fig. \ref{figs:NNS_result}).

Remarkably, the seven groups of ALMA compact cores roughly correspond to 
dense clumps visible as closed contours in the ArT$\acute{\rm e}$MiS 350 $\mu$m map 
of the NGC~6334 filament at 8$\arcsec$ resolution (see Figs. \ref{figs}c and \ref{figs:n2h+_blue}). 
For the purpose of this paper, we used GAUSSCLUMPS \citep{Stutzki90} to characterize the properties of the ArT$\acute{\rm e}$MiS clumps. 
We applied GAUSSCLUMPS with a detection threshold 
of 1 Jy/8\arcsec -beam corresponding to 5$\sigma$ (where 1$\sigma$=0.2 Jy/8\arcsec -beam). 
A total of seven clumps were identified in this way within the footprint of the main filament, 
defined as the intersection of the area within $\pm$30$\arcsec$ (or $\pm$0.25\,pc)\footnote{The area within $\pm$0.25\,pc of the filament crest 
extends well into the power-law wing of the filament radial profile (see Fig.~3a of \citealp{Andre16}) and 
includes the bulk of the ArT\'eMiS 350 $\mu$m emission from the filament  (see Fig.~\ref{figs:NNS_result}).}
from the filament crest and the interior of the 5 Jy/8\arcsec -beam ArT$\acute{\rm e}$MiS contour  
(see solid white contour in Fig.~\ref{figs:NNS_result}), 
whose positions, sizes, and estimated masses are given in Table \ref{table:ArTeMiS_clumps}. 
These seven ArT$\acute{\rm e}$MiS clumps are also identified with the \textsl{getsources} \citep{Menshchikov12,Menshchikov13} and REINHLOD \citep{Berry07} 
source extraction algorithms, as described in Appendix~\ref{Sect:comparison_IDs}.
They correspond to six groups of ALMA cores (groups 1, 2, 3, 5, 6, 7).
More precisely, if we consider a group of ALMA dense cores to be associated with an ArT$\acute{\rm e}$MiS clump 
when all of the cores in the group lie within the closed contours of the clump (Table \ref{table:ArTeMiS_clumps}), 
then we find that each ArT$\acute{\rm e}$MiS clump consists of at least two ALMA cores (Fig.~\ref{figs:ArTeMiS_ALMA_core_association}a).

To test whether the observed clustering of ALMA cores and close association with ArT$\acute{\rm e}$MiS clumps
may also be present in the case of randomly distributed cores, we inserted 26 sources at random positions within the footprint of the main filament (see solid white contour in Fig.~\ref{figs:NNS_result}) using the python module {\it random} and then applied the same NNS analysis. A total of 100 realizations of such random source distributions were constructed. Two examples of resulting NNS dendrogram trees are displayed in Fig.~\ref{figs_NNStest}, where they are compared to the NNS tree obtained for the real ALMA cores. It can be seen that randomly distributed sources tend to be less clustered than the observed cores, that is, a higher number of randomly-placed sources are isolated (not grouped by the NNS analysis) compared to the observations. In contrast to the observed cores, there is also no clear association between the groups of randomly-placed sources and the ArT$\acute{\rm e}$MiS clumps. The clustering of the observed ALMA cores within the ArT$\acute{\rm e}$MiS clumps is highly significant, at the $> 5\sigma$ level according to binomial statistics: 
16 out of 26 cores ($> 60\%$) lie within the FWHM ellipses of the clumps, while only $3\pm2$ ($\sim 13\%$) of 26 randomly-placed objects would be expected within the clumps. Furthermore, the mean separation between observed cores ($0.04\pm 0.01$ pc) is smaller than the mean separation between randomly-placed sources found over 100 realizations ($0.06\pm 0.01$ pc).

It is also apparent in the top row of Fig.~\ref{figs:NNS_result} that the observed cores lie significantly closer 
to the filament crest than randomly-placed objects within the filament footprint (solid white contour). Quantitatively, the distribution of offsets between the observed cores and the filament crest has a median value of only 0.02 pc (lower quartile: 0.01 pc, upper quartile: 0.03 pc) while the median offset between randomly-placed sources and the crest is 0.07 pc (lower quartile: 0.03 pc, upper quartile: 0.13 pc). In other words, the spatial distribution of observed cores is almost one-dimensional (1D) along the filament crest. 
To further test whether significant 1D grouping of the cores exists along the crest, we repeated the same experiment with randomly-placed objects as described above but using a narrower footprint, defined as the area within $\pm$6$\arcsec$ (or $0.05\,$pc) of the filament crest  (dashed white contour in Fig.~\ref{figs:NNS_result}). 
The transverse width of this inner filament footprint is $\sim\,$0.1\,pc, which is comparable to the dispersion of observed core positions about the crest. In this case, the mean separation between randomly-placed sources becomes identical to the mean separation between observed cores ($0.04\pm 0.01\,$pc). 
The observed ALMA cores nevertheless remain more closely associated with the ArT$\acute{\rm e}$MiS clumps than randomly-distributed objects along the filament crest: only $8\pm3$ of 26 randomly-placed sources would be expected within the clumps, while 16 are observed, a difference which is significant at the $\sim 3\, \sigma$ level according to binomial statistics. We conclude that, even in 1D, there is marginal evidence of non-random core grouping along the filament axis.

The projected nearest-neighbor separation between the ArT$\acute{\rm e}$MiS clumps ranges from 0.2 pc to 0.3 pc, 
and the typical projected separation between ALMA dense cores embedded within a given ArT$\acute{\rm e}$MiS clump is  $0.04\pm0.01$ pc. These observational findings are interesting to compare with theoretical expectations. When the line mass of a cylindrical gas filament is close to the critical mass per unit length (thermal case) or the virial mass per unit length (nonthermal case, more appropriate here -- see end of Sect.~\ref{intro}), the filament is expected to fragment into overdensities with a typical separation of about four times the filament width according to self-similar solutions which describe the collapse of an isothermal filament under the effect of self-gravity \citep[e.g][]{Inutsuka92, Inutsuka97}. With a typical inner width of $0.15\pm 0.05\,$pc from the ArT\'eMiS results \citep{Andre16}, the NGC~6334 filament can therefore 
be expected to fragment with a characteristic separation of $\sim \, 0.6\pm 0.2$ pc. This separation scale is roughly consistent with the projected separation of 0.2--0.3 pc observed between ArT$\acute{\rm e}$MiS clumps, assuming a plausible inclination angle of $\sim 30^\circ$ between the filament axis and the line of sight.

\begin{figure}
\centering
\includegraphics[width=90mm, angle=0]{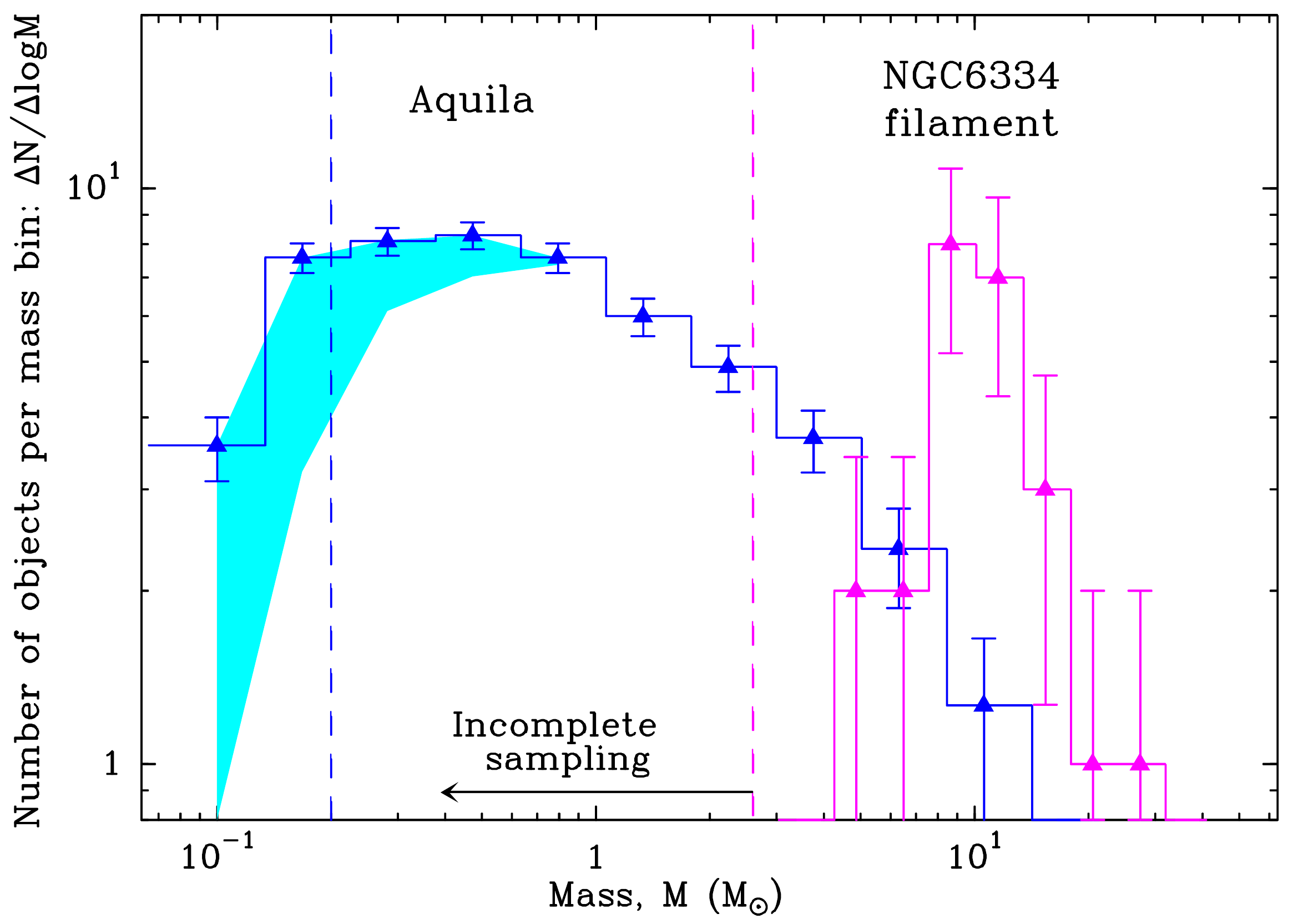}
\caption{Mass distribution of the 26 ALMA cores identified in the NGC~6334 main filament (magenta) 
compared to a scaled version of the prestellar CMF observed in the Aquila cloud \citep[blue, from][]{Konyves15}. 
In both NGC~6334 and Aquila, cores were extracted from the data using the same algorithm \textsl{getsources}.
For easier comparison, the Aquila CMF was re-normalized to have a peak value comparable to that of the NGC~6334 CMF. 
All error bars correspond to $\sqrt{N}$ counting statistics. 
The magenta vertical dashed line indicates the estimated 90\% completeness level 
($2.6\, M_{\odot}$) of our ALMA 3.2\,mm census of dense cores 
in the NGC~6334 filament. 
The blue vertical dashed line marks the 90\% 
completeness level of the Aquila CMF (0.2$M_{\odot}$). 
(The blue shaded area reflects uncertainties associated 
with the uncertain classification of starless cores as bound or unbound objects.)
}
\label{figs:CMF}
\end{figure}

On the other hand, the effective Jeans length $\lambda_{\rm J,eff}$ or Bonnor-Ebert diameter $D_{\rm BE,eff}$ within the filament 
and its clumps may be estimated as \citep[cf.][]{Bonnor56}:

\begin{equation}
\lambda_{\rm J,eff} \equiv D_{\rm BE,eff} \approx 0.98 \sqrt{\frac{C_{\rm s,eff}^2}{G\, \rho_{\rm clump}}}, 
\end{equation}

\noindent where 

\begin{equation}
C_{\rm s,eff}= \sqrt{\sigma_{\rm N_2H^+}^2 + kT_{\rm k} (\frac{1}{m}-\frac{1}{m_{\rm N_2H^+}})}, 
\end{equation}

\noindent and $G$, $\rho_{\rm clump}$, $\sigma_{\rm N_2H^+}$, $T_{\rm k}$, $m$, and $m_{\rm N_2H^+}$  are the gravitational constant, the density of each ArT$\acute{\rm e}$MiS clump (see Table \ref{table:ArTeMiS_clumps}), 
the velocity dispersion measured in N$_2$H$^+$, the gas kinetic temperature, the mean molecular mass, and the mass of the N$_2$H$^+$ molecule, respectively. Assuming a gas kinetic temperature of 20 K, which corresponds to the median dust temperature derived from {\it Herschel} data along the crest of the filament \citep{Andre16,Tige17} 
and using the velocity dispersion measured in N$_2$H$^+$ for each ArT$\acute{\rm e}$MiS clump ($\sigma_{\rm N_2H^+}$= $\delta V_{\rm FWHM}$/$\sqrt{8\ln(2)}$, where $\delta V_{\rm FWHM}$ is the FWHM linewidth), the effective Jeans length in the clumps of the filament is estimated to range from $\sim$0.04 pc to $\sim$0.3 pc (median: 0.08 pc). This is roughly consistent with the typical projected separation between ALMA dense cores (0.04$\pm$0.01 pc) assuming random projection effects within the ArT$\acute{\rm e}$MiS clumps. 

These two characteristic separation scales are suggestive of two distinct fragmentation modes within the NGC~6334 filament: (i) a cylindrical fragmentation mode into clumps or groups of cores with a separation of $\sim 4$ times the filament width, and (ii) a spherical Jeans-like fragmentation mode into compact cores 
with a separation on the order of the effective Jeans length. 
Similar bimodal fragmentation patterns were first reported by \citet{Takahashi13} in Orion OMC3 
and \citet{Kainulainen13} in the infrared dark cloud (IRDC) G11.11-0.12 (see also \citealp{Teixeira16} and \citealp{Kainulainen17}). 
Recent theoretical works on filament fragmentation have tried to account for these two fragmentation modes 
based on perturbations to standard cylinder fragmentation models \citep[cf.][]{Clarke16, Gritschneder17, Lee17}.

\subsection{Unusually massive cores in the NGC 6334 filament}\label{cmf}

Remarkably, the median mass of the ALMA cores ($9.6^{+3.0}_{-1.9}\, M_{\odot}$) is an order of magnitude higher 
than the peak of the prestellar core mass function (CMF) at $\sim 0.6\, M_\odot $ as measured with {\it Herschel} 
in nearby clouds \citep[e.g][]{Konyves15}. This is illustrated in Fig.~\ref{figs:CMF} which compares a rough estimate of the CMF in the NGC~6334 filament based on the present ALMA study to the CMF derived from {\it Herschel} data in the Aquila cloud. 
Although the two data sets differ somewhat in nature and wavelength (e.g., ALMA 3.1~mm vs. {\it Herschel} 160--500$\, \mu$m), 
cores were extracted from dust continuum maps using the same algorithm (\textsl{getsources}) in both cases. 
Moreover, the ALMA 3.1~mm continuum data used here are sensitive to the typical size scales of dense cores identified 
with {\it Herschel} in Gould Belt clouds ($\sim$\,0.02~pc to $\sim$\,0.1~pc -- see Sect.~\ref{ALMA_obs} and Fig.~7 of \citealp{Konyves15}). 
The two CMFs shown in Fig.~\ref{figs:CMF} are thus directly comparable.
The difference in typical core mass between NGC~6334 and Aquila largely exceeds possible uncertainties in the dust emissivity at 3.1~mm. 
For instance, if the dust emissivity index $\beta = 1.5$ (instead of  $\beta = 2$ as assumed in Table~\ref{table:cont_getsources_12m_only} and Fig.~\ref{figs:CMF}), the median core mass in NGC~6334 would still be a factor $\sim$5 higher than the CMF peak in Aquila. The median mass of observed ALMA cores essentially corresponds to the peak of the NGC~6334 CMF, and lies more than a factor of three above the estimated 90\% completeness level ($2.6\, M_{\odot}$) of the present core census\footnote{The 90\% completeness level was estimated to be $2.6\, M_{\odot}$ assuming Gaussian statistics and a 5$\sigma$ peak-flux detection threshold of 0.7~mJy/beam. Under these assumptions, $\sim$10\% of sources with intrinsic peak fluxes at the 6.3$\sigma$ level of $\sim 0.9$~mJy/beam will have measured peak fluxes below 5$\sigma$. Thus, one can expect the 90\% completeness level to roughly correspond to the 6.3$\sigma$ level \citep[cf.][]{Belloche11}.}. While the NGC~6334 CMF shown in Fig.~\ref{figs:CMF} is admittedly quite uncertain due to, for instance, low-number statistics and uncertainties in the 3.1~mm dust emissivity, 
we stress that it represents one of the first estimates of the CMF generated by a single, 
massive filament\footnote{Despite the presence of fiber-like sub-structures, we regard the prominent filamentary structure seen in the ArT\'eMiS 350\,$\mu$m continuum map and the ALMA N$_2$H$^+$(1--0) integrated intensity map (cf. Fig.~\ref{figs}) as a single filament, comparable to the many filaments detected with {\it Herschel} in nearby clouds \citep{Arzoumanian11, Arzoumanian18} and to the B211/B213 filament in Taurus \citep[][and Sect.~\ref{fibers}]{Palmeirim13,Hacar13}.} (see \citealp[][]{Takahashi13, Zhang15, Ohashi16} for examples of CMFs in somewhat less massive filaments or IRDCs with $M_{line} < 500\, M_\odot $/pc). 

The median core mass derived from the ALMA data is roughly consistent with the 
effective critical Bonnor-Ebert mass $M_{\rm BE,eff}$ in the clumps of the filament,  
which can be expressed as  \citep{Bonnor56}:

\begin{equation}
M_{\rm BE,eff} = 1.18 \frac{C_{\rm s,eff}^3}{\sqrt{G^3\rho_{\rm clump}}},
\end{equation}

\noindent 
or about 7.2 $M_{\odot}$ (lower quartile: 5.0 $M_{\odot}$, upper quartile: 23.9 $M_{\odot}$). 
Together with the results of Sect.~\ref{two_levels} on core grouping, this 
supports the view that the ALMA dense cores arise from effective Jeans-like fragmentation of the ArT$\acute{\rm e}$MiS clumps. 
Our results also suggest that more massive cores may form in denser/more massive filaments and
are consistent with a picture in which the global prestellar CMF, and possibly the stellar initial mass function (IMF) itself, 
originate from the superposition of the CMFs generated by individual filaments with a whole spectrum of masses per unit length \citep[][]{Andre+2019}.

\subsection{Likelihood of the NGC 6334 filament being a system of two velocity-coherent fibers}\label{fibers}

As described in Sect. \ref{Sect:Filament_ID}, our ALMA observations show that the $\sim \,$0.15-pc wide filament detected in the ArT$\acute{\rm e}$MiS 350 $\mu$m 
dust continuum map is sub-structured in five fiber-like N$_2$H$^+$(1--0) components with different velocities. 
The typical length of these fiber-like sub-structures is 0.5$\pm$0.4 pc and the typical velocity difference between them is 0.8 km s$^{-1}$  
(standard deviation of the five N$_2$H$^+$ velocity components). 
The N$_2$H$^+$ velocity-coherent sub-structures may be broadly categorized into two groups. 
The first group consists of the N$_2$H$^+$ fiber-like sub-structures  F-1 and F-2, which are also detected in other molecular line tracers 
of dense gas such as CH$_3$CCH, H$_2$CS, HC$_3$N, and HC$_5$N 
and harbor compact dense cores detected in 3.1~mm continuum emission.
The second group  (F-3, F-4, and F-5) is made up of velocity-coherent sub-structures detected only in N$_2$H$^+$ 
and which do not seem to contain dense cores. 
There are at least two possible explanations for these two groups. 
First, the 
physical excitation conditions in the two groups of 
sub-structures may differ.  
Second, the low velocity resolution of the current CH$_3$CCH, H$_2$CS, HC$_3$N line data (see Table \ref{table:alma_obs}) may prevent the identification of fiber-like features in these tracers. 
To test the former explanation,
observations in several transitions of each species would be required to estimate the excitation temperature, 
column density, and chemical abundance of the various molecules.
To investigate the latter effect,  ALMA observations of the same tracers at higher spectral resolution would be required. 

The presence of velocity-coherent fiber-like sub-structures in molecular filaments was first reported by \citet{Hacar13} in the case of the low-mass star-forming filament  B211--B213 in Taurus ($d\sim$140 pc). In this filament, Hacar et al. used their friends in velocity (FIVE) algorithm to identify at least 20  velocity-coherent components in N$_2$H$^+$ and C$^{18}$O (see Table 3 in \citealp{Hacar13}), which were subsequently called fibers. 
Since then, similar velocity-coherent components have also been detected in N$_2$H$^+$ in the IRDC 
G035.39-00.33 \citep{Henshaw14}, the NGC~1333 protocluster \citep{Hacar17}, IRDC G034.43+00.24 \citep{Barnes18}, and the Orion A integral-shaped filament \citep{Hacar18}.
The fiber-like sub-structures identified in NGC~1333 and Orion~A may, however, differ in nature from those observed in the Taurus B211/B213 filament and in the present target NGC~6334 \citep[see also ][]{Clarke17}. 
The velocity-coherent  sub-structures observed in the NGC~1333 and Orion~A cases are indeed well separated in the plane of the sky, while those in the Taurus and NGC~6334 filaments overlap in the sky 
and can mostly be distinguished in PPV space. 

The typical length (0.6$\pm$0.5 pc) and velocity difference between components (0.7 km s$^{-1}$) 
reported by  \citet{Hacar13} for the fiber-like sub-structures of the (low-mass) B211/B213 filament 
in Taurus are remarkably similar to the properties estimated here for the velocity-coherent sub-structures 
of  the (high-mass) NGC~6334 filament. We further note that \citet{Hacar13} divided up their B211/B213 fibers into two groups, fertile and sterile, depending on whether they contained dense cores or not. Most of the 35 velocity-coherent sub-structures identified in  B211/B213 were sterile and detected mostly in C$^{18}$O(1--0), while only 7 fiber-like sub-structures were fertile and also detected in N$_2$H$^+$(1--0). This is reminiscent of the situation found here for the NGC~6334 filament, where the two sub-structures detected in multiple dense gas tracers (F-1 and F-2) are the only two fertile fibers harboring ALMA dense cores. We argue below that these two categories of fiber-like structures differ in physical nature and possibly origin.

\subsection{Possible origin(s) of the fiber-like sub-structures}

Three scenarios for the formation of velocity-coherent fiber-like sub-structures have been proposed in the literature.
One is the ``fray and fragment'' scenario proposed by \citet{Tafalla15} and supported by \citet{Clarke17}. 
In this scenario, a main filament forms first by collision of two supersonic turbulent gas flows. 
Then, the main filament fragments into an intertwined system of velocity-coherent sub-structures, due to a combination of residual turbulent motions and self-gravity. 
In this picture, the sub-structures are formed by fragmentation of a single filament, and the velocity-coherent sub-structures 
are expected to be roughly aligned with the main filament \citep[cf.][]{Smith14}.
The second scenario is the  ``fray and gather'' scenario, in which turbulent compression first generates short, velocity-coherent filamentary structures within the parent cloud, 
which are then gathered by large-scale collapse of the cloud, as proposed by \citet{Smith14}. 
In a third, alternative picture, a dense star-forming filament forms within a shell-like molecular gas layer 
as a result of large-scale anisotropic compression associated with, for instance, expanding bubble(s) or cloud-cloud collision \citep[][]{Chen14,Inutsuka15,Inoue18}, 
and subsequently grows in mass by accreting ambient gas from the surrounding shell-like structure due to its own gravitational potential \citep[cf.][]{Palmeirim13,Shimajiri18}. 
This accretion process supplies gravitational energy to the dense filament, which is then converted into turbulent kinetic energy in the form of MHD waves \citep{Hennebelle13}, 
explaining the increase in velocity dispersion with column density observed for thermally supercritical filaments \citep{Arzoumanian13}. 
A quasi-stationary state is reached as a result of a dynamical equilibrium between accretion-driven MHD turbulence 
and the dissipation of this MHD turbulence owing to ion-neutral friction, possibly accounting for the roughly constant filament width of $\sim 0.1\, $pc \citep{Hennebelle13}. 
In this ``compress, accrete, and fragment'' scenario, sterile fiber-like structures would correspond to portions of the accretion flow  
onto the central filament \citep[see also][]{Clarke18}, 
while fertile fiber-like structures would be the direct imprint of accretion-driven MHD waves within the main filament system.

While further observational constraints will be needed to fully discriminate between the above three pictures, the available constraints tend to support a scenario intermediate
between the fray and fragment and the fray and gather picture, perhaps more similar to the compress, accrete, and fragment picture that we propose here. 
Indeed, the NGC~6334 region may be affected by cloud-cloud collision \citep[][see also Appendix~\ref{Sect:origin_high_blue}]{Fukui18},  and 
some of the N$_2$H$^+$ velocity-coherent sub-structures identified here (e.g., F-5) are not aligned with, but roughly perpendicular to,  
the main filament traced at 350$\, \mu$m by ArT\'eMiS and in N$_2$H$^+$ by the two main sub-structures  (F-1, F-2  in Table \ref{table:N$_2$H$^+$_filament}). 
Furthermore, \citet{Shimajiri18} recently reported  kinematic evidence that the B211/B213 filament in Taurus may have formed inside a shell-structure 
resulting from large-scale compression. 

\section{Conclusions}\label{Sect:Conclusions}

To study the detailed density and velocity structure of the massive filament in the NGC~6334 complex, 
we carried out ALMA observations at $\sim\,$$3\arcsec$ resolution in the 3.1 mm continuum and 
the N$_2$H$^+$(1-0), HC$_5$N(36-35), HNC(1-0), HC$_3$N(10-9), CH$_3$CCH(6-5), H$_2$CS(3-2) lines.
Our main results may be summarized as follows: 

\begin{enumerate}

\item Both the 3.1 mm continuum emission and the N$_2$H$^+$(1--0), HC$_5$N(36--35), HC$_3$N(10--9), CH$_3$CCH(6--5),  H$_2$CS(3--2) lines detected with ALMA trace the dusty filament imaged earlier at 350 $\mu$m at lower ($8\arcsec $) resolution with the ArT$\acute{\rm e}$MiS  camera on APEX. 

\item We identified a total of 40 compact dense cores in the ALMA 3.1 mm continuum map, 26 of them being embedded in the NGC~6334 filament. The majority (21/26 or 80\%) of these dense cores are also detected in N$_2$H$^+$(1--0) emission. The median core mass is $10^{+3}_{-2}\, M_\odot $ 
(lower quartile: 8 $M_{\odot}$, upper quartile: 13 $M_{\odot}$), compared to a 5$\sigma$ mass sensitivity of $2\, M_{\odot}$. 

\item The CMF derived from the sample of ALMA cores in the NGC~6334 filament (Fig.~\ref{figs:CMF}) presents a peak at the median core mass of $\sim$10\,$M_\odot $, which lies an order of magnitude higher than the peak of 
the prestellar CMF measured with {\it Herschel} in nearby clouds.

\item The projected separation between ALMA dense cores is 0.03--0.1 pc, 
which is roughly consistent with the effective Jeans length within the filament. 
The ALMA cores can be grouped into seven groups, approximately corresponding 
to dense clumps seen in the ArT$\acute{\rm e}$MiS 350 $\mu$m continuum map.
The projected separation between these groups is 0.2--0.3 pc, which roughly agrees 
with the characteristic spacing of four times the filament width 
expected from the linear fragmentation theory of nearly isothermal gas cylinders. 
These two distinct fragmentation scales are suggestive of two fragmentation modes:
a cylindrical mode corresponding to groups of cores, and 
a spherical, Jeans-like mode corresponding to cores within groups.

\item We also identified five fiber-like, velocity-coherent sub-structures within the filament
by applying the \textsl{getfilaments} algorithm to the ALMA N$_2$H$^+$(1--0) data cube.
The typical length of these fiber-like structures is 0.5 pc and the projected velocity difference between them
is $\sim \,$0.8 km s$^{-1}$. 
Only two or three of these five velocity-coherent features are well aligned with the NGC~6334 filament 
and may represent genuine, intertwined fiber sub-structures. The other two detected velocity-coherent features 
may rather trace accretion flows onto the main filament.

\item 
With the important exception of the typical core mass (which is here an order of magnitude higher), 
the fragmentation properties and velocity structure of the massive ($\ga 500\, M_\odot /$pc) filament in NGC~6334 are remarkably 
similar to the properties observed by \citet{Hacar13} and \citet{Tafalla15} for the low-mass  ($\sim 50\, M_\odot /$pc) B211/B213 filament 
in the Taurus cloud.

\item As both regions appear to be affected by large-scale compressive flows, 
we suggest that the density and velocity sub-structure observed in the NGC~6334 and the Taurus filament 
may have originated through a similar mechanism, which we dub ``compress, accrete, and fragment''.

\end{enumerate}

\begin{acknowledgements}
 This paper makes use of the following ALMA data: ADS/JAO.ALMA\#2015.1.01404.S. ALMA is a partnership of ESO (representing its member states), NSF (USA) and NINS (Japan), 
 together with NRC (Canada),NSC and ASIAA (Taiwan), 
 and KASI (Republic of Korea), in cooperation with the Republic of Chile. The Joint ALMA Observatory is operated by ESO, AUI/NRAO and NAOJ.
 This work was supported by the ANR-11-BS56-010 project ``STARFICH" and the European Research Council under the European Union's Seventh Framework Programme 
 (ERC Advanced Grant Agreement no. 291294 --  `ORISTARS'). YS also received support from the ANR project NIKA2SKY (grant agreement ANR-15-CE31-0017). 
  This work was supported by NAOJ ALMA Scientific Research Grant
Numbers 2017-04A.
 We also acknowledge support from the French national programs of CNRS/INSU on stellar and ISM physics (PNPS and PCMI). 
\end{acknowledgements}

\bibliographystyle{aa}
\bibliography{ALMA_NGC6334_final.bbl}

\begin{thebibliography}{69}
\expandafter\ifx\csname natexlab\endcsname\relax\def\natexlab#1{#1}\fi

\bibitem[{{Andr{\'e}}(2017)}]{Andre17}
{Andr{\'e}}, P. 2017, Comptes Rendus Geoscience, 349, 187

\bibitem[{{Andr{\'e}} {et~al.}(2019){Andr{\'e}}, {Arzoumanian}, {K{\"o}nyves},
  {Shimajiri}, \& {Palmeirim}}]{Andre+2019}
{Andr{\'e}}, P., {Arzoumanian}, D., {K{\"o}nyves}, V., {Shimajiri}, Y., \&
  {Palmeirim}, P. 2019, \aap, 629, L4

\bibitem[{{Andr{\'e}} {et~al.}(2014){Andr{\'e}}, {Di Francesco},
  {Ward-Thompson}, {Inutsuka}, {Pudritz}, \& {Pineda}}]{Andre14}
{Andr{\'e}}, P., {Di Francesco}, J., {Ward-Thompson}, D., {et~al.} 2014,
  Protostars and Planets VI, 27

\bibitem[{{Andr{\'e}} {et~al.}(2010){Andr{\'e}}, {Men'shchikov}, {Bontemps},
  {K{\"o}nyves}, {Motte}, {Schneider}, {Didelon}, {Minier}, {Saraceno},
  {Ward-Thompson}, {di Francesco}, {White}, {Molinari}, {Testi}, {Abergel},
  {Griffin}, {Henning}, {Royer}, {Mer{\'{\i}}n}, {Vavrek}, {Attard},
  {Arzoumanian}, {Wilson}, {Ade}, {Aussel}, {Baluteau}, {Benedettini},
  {Bernard}, {Blommaert}, {Cambr{\'e}sy}, {Cox}, {di Giorgio}, {Hargrave},
  {Hennemann}, {Huang}, {Kirk}, {Krause}, {Launhardt}, {Leeks}, {Le Pennec},
  {Li}, {Martin}, {Maury}, {Olofsson}, {Omont}, {Peretto}, {Pezzuto}, {Prusti},
  {Roussel}, {Russeil}, {Sauvage}, {Sibthorpe}, {Sicilia-Aguilar}, {Spinoglio},
  {Waelkens}, {Woodcraft}, \& {Zavagno}}]{Andre10}
{Andr{\'e}}, P., {Men'shchikov}, A., {Bontemps}, S., {et~al.} 2010, \aap, 518,
  L102

\bibitem[{{Andr{\'e}} {et~al.}(2016){Andr{\'e}}, {Rev{\'e}ret}, {K{\"o}nyves},
  {Arzoumanian}, {Tig{\'e}}, {Gallais}, {Roussel}, {Le Pennec}, {Rodriguez},
  {Doumayrou}, {Dubreuil}, {Lortholary}, {Martignac}, {Talvard}, {Delisle},
  {Visticot}, {Dumaye}, {De Breuck}, {Shimajiri}, {Motte}, {Bontemps},
  {Hennemann}, {Zavagno}, {Russeil}, {Schneider}, {Palmeirim}, {Peretto},
  {Hill}, {Minier}, {Roy}, \& {Rygl}}]{Andre16}
{Andr{\'e}}, P., {Rev{\'e}ret}, V., {K{\"o}nyves}, V., {et~al.} 2016, \aap,
  592, A54

\bibitem[{{Arzoumanian} {et~al.}(2011){Arzoumanian}, {Andr{\'e}}, {Didelon},
  {K{\"o}nyves}, {Schneider}, {Men'shchikov}, {Sousbie}, {Zavagno}, {Bontemps},
  {di Francesco}, {Griffin}, {Hennemann}, {Hill}, {Kirk}, {Martin}, {Minier},
  {Molinari}, {Motte}, {Peretto}, {Pezzuto}, {Spinoglio}, {Ward-Thompson},
  {White}, \& {Wilson}}]{Arzoumanian11}
{Arzoumanian}, D., {Andr{\'e}}, P., {Didelon}, P., {et~al.} 2011, \aap, 529, L6

\bibitem[{{Arzoumanian} {et~al.}(2019){Arzoumanian}, {Andr{\'e}},
  {K{\"o}nyves}, {Palmeirim}, {Roy}, {Schneider}, {Benedettini}, {Didelon}, {Di
  Francesco}, {Kirk}, \& {Ladjelate}}]{Arzoumanian18}
{Arzoumanian}, D., {Andr{\'e}}, P., {K{\"o}nyves}, V., {et~al.} 2019, \aap,
  621, A42

\bibitem[{{Arzoumanian} {et~al.}(2013){Arzoumanian}, {Andr{\'e}}, {Peretto}, \&
  {K{\"o}nyves}}]{Arzoumanian13}
{Arzoumanian}, D., {Andr{\'e}}, P., {Peretto}, N., \& {K{\"o}nyves}, V. 2013,
  \aap, 553, A119

\bibitem[{{Barnes} {et~al.}(2018){Barnes}, {Henshaw}, {Caselli},
  {Jim{\'e}nez-Serra}, {Tan}, {Fontani}, {Pon}, \& {Ragan}}]{Barnes18}
{Barnes}, A.~T., {Henshaw}, J.~D., {Caselli}, P., {et~al.} 2018, \mnras, 475,
  5268

\bibitem[{{Belloche} {et~al.}(2011){Belloche}, {Schuller}, {Parise},
  {Andr{\'e}}, {Hatchell}, {J{\o}rgensen}, {Bontemps}, {Wei{\ss}}, {Menten}, \&
  {Muders}}]{Belloche11}
{Belloche}, A., {Schuller}, F., {Parise}, B., {et~al.} 2011, \aap, 527, A145

\bibitem[{{Berry} {et~al.}(2007){Berry}, {Reinhold}, {Jenness}, \&
  {Economou}}]{Berry07}
{Berry}, D.~S., {Reinhold}, K., {Jenness}, T., \& {Economou}, F. 2007, in
  Astronomical Society of the Pacific Conference Series, Vol. 376, Astronomical
  Data Analysis Software and Systems XVI, ed. R.~A. {Shaw}, F.~{Hill}, \& D.~J.
  {Bell}, 425

\bibitem[{{Bonnor}(1956)}]{Bonnor56}
{Bonnor}, W.~B. 1956, \mnras, 116, 351

\bibitem[{{Chen} \& {Ostriker}(2014)}]{Chen14}
{Chen}, C.-Y. \& {Ostriker}, E.~C. 2014, \apj, 785, 69

\bibitem[{{Clarke} {et~al.}(2017){Clarke}, {Whitworth}, {Duarte-Cabral}, \&
  {Hubber}}]{Clarke17}
{Clarke}, S.~D., {Whitworth}, A.~P., {Duarte-Cabral}, A., \& {Hubber}, D.~A.
  2017, \mnras, 468, 2489

\bibitem[{{Clarke} {et~al.}(2016){Clarke}, {Whitworth}, \& {Hubber}}]{Clarke16}
{Clarke}, S.~D., {Whitworth}, A.~P., \& {Hubber}, D.~A. 2016, \mnras, 458, 319

\bibitem[{{Clarke} {et~al.}(2018){Clarke}, {Whitworth}, {Spowage},
  {Duarte-Cabral}, {Suri}, {Jaffa}, {Walch}, \& {Clark}}]{Clarke18}
{Clarke}, S.~D., {Whitworth}, A.~P., {Spowage}, R.~L., {et~al.} 2018, \mnras,
  479, 1722

\bibitem[{{Fiege} \& {Pudritz}(2000)}]{Fiege00}
{Fiege}, J.~D. \& {Pudritz}, R.~E. 2000, \mnras, 311, 85

\bibitem[{{Fukui} {et~al.}(2015){Fukui}, {Harada}, {Tokuda}, {Morioka},
  {Onishi}, {Torii}, {Ohama}, {Hattori}, {Nayak}, {Meixner}, {Sewi{\l}o},
  {Indebetouw}, {Kawamura}, {Saigo}, {Yamamoto}, {Tachihara}, {Minamidani},
  {Inoue}, {Madden}, {Galametz}, {Lebouteiller}, {Mizuno}, \& {Chen}}]{Fukui15}
{Fukui}, Y., {Harada}, R., {Tokuda}, K., {et~al.} 2015, \apjl, 807, L4

\bibitem[{{Fukui} {et~al.}(2018){Fukui}, {Kohno}, {Yokoyama}, {Torii},
  {Hattori}, {Sano}, {Nishimura}, {Ohama}, {Yamamoto}, \&
  {Tachihara}}]{Fukui18}
{Fukui}, Y., {Kohno}, M., {Yokoyama}, K., {et~al.} 2018, \pasj, 70, S41

\bibitem[{{Gao} \& {Solomon}(2004)}]{Gao04}
{Gao}, Y. \& {Solomon}, P.~M. 2004, \apj, 606, 271

\bibitem[{{Gritschneder} {et~al.}(2017){Gritschneder}, {Heigl}, \&
  {Burkert}}]{Gritschneder17}
{Gritschneder}, M., {Heigl}, S., \& {Burkert}, A. 2017, \apj, 834, 202

\bibitem[{{Hacar} {et~al.}(2017){Hacar}, {Tafalla}, \& {Alves}}]{Hacar17}
{Hacar}, A., {Tafalla}, M., \& {Alves}, J. 2017, \aap, 606, A123

\bibitem[{{Hacar} {et~al.}(2018){Hacar}, {Tafalla}, {Forbrich}, {Alves},
  {Meingast}, {Grossschedl}, \& {Teixeira}}]{Hacar18}
{Hacar}, A., {Tafalla}, M., {Forbrich}, J., {et~al.} 2018, \aap, 610, A77

\bibitem[{{Hacar} {et~al.}(2013){Hacar}, {Tafalla}, {Kauffmann}, \&
  {Kov{\'a}cs}}]{Hacar13}
{Hacar}, A., {Tafalla}, M., {Kauffmann}, J., \& {Kov{\'a}cs}, A. 2013, \aap,
  554, A55

\bibitem[{{Hennebelle} \& {Andr{\'e}}(2013)}]{Hennebelle13}
{Hennebelle}, P. \& {Andr{\'e}}, P. 2013, \aap, 560, A68

\bibitem[{{Henshaw} {et~al.}(2014){Henshaw}, {Caselli}, {Fontani},
  {Jim{\'e}nez-Serra}, \& {Tan}}]{Henshaw14}
{Henshaw}, J.~D., {Caselli}, P., {Fontani}, F., {Jim{\'e}nez-Serra}, I., \&
  {Tan}, J.~C. 2014, \mnras, 440, 2860

\bibitem[{{Hildebrand}(1983)}]{Hildebrand83}
{Hildebrand}, R.~H. 1983, \qjras, 24, 267

\bibitem[{{Hill} {et~al.}(2011){Hill}, {Motte}, {Didelon}, {Bontemps},
  {Minier}, {Hennemann}, {Schneider}, {Andr{\'e}}, {Men'shchikov}, {Anderson},
  {Arzoumanian}, {Bernard}, {di Francesco}, {Elia}, {Giannini}, {Griffin},
  {K{\"o}nyves}, {Kirk}, {Marston}, {Martin}, {Molinari}, {Nguyen Luong},
  {Peretto}, {Pezzuto}, {Roussel}, {Sauvage}, {Sousbie}, {Testi},
  {Ward-Thompson}, {White}, {Wilson}, \& {Zavagno}}]{Hill11}
{Hill}, T., {Motte}, F., {Didelon}, P., {et~al.} 2011, \aap, 533, A94

\bibitem[{{Inoue} {et~al.}(2018){Inoue}, {Hennebelle}, {Fukui}, {Matsumoto},
  {Iwasaki}, \& {Inutsuka}}]{Inoue18}
{Inoue}, T., {Hennebelle}, P., {Fukui}, Y., {et~al.} 2018, \pasj, 70, S53

\bibitem[{{Inutsuka} {et~al.}(2015){Inutsuka}, {Inoue}, {Iwasaki}, \&
  {Hosokawa}}]{Inutsuka15}
{Inutsuka}, S.-i., {Inoue}, T., {Iwasaki}, K., \& {Hosokawa}, T. 2015, \aap,
  580, A49

\bibitem[{{Inutsuka} \& {Miyama}(1992)}]{Inutsuka92}
{Inutsuka}, S.-I. \& {Miyama}, S.~M. 1992, \apj, 388, 392

\bibitem[{{Inutsuka} \& {Miyama}(1997)}]{Inutsuka97}
{Inutsuka}, S.-I. \& {Miyama}, S.~M. 1997, \apj, 480, 681

\bibitem[{{Kainulainen} {et~al.}(2013){Kainulainen}, {Ragan}, {Henning}, \&
  {Stutz}}]{Kainulainen13}
{Kainulainen}, J., {Ragan}, S.~E., {Henning}, T., \& {Stutz}, A. 2013, \aap,
  557, A120

\bibitem[{{Kainulainen} {et~al.}(2017){Kainulainen}, {Stutz}, {Stanke},
  {Abreu-Vicente}, {Beuther}, {Henning}, {Johnston}, \&
  {Megeath}}]{Kainulainen17}
{Kainulainen}, J., {Stutz}, A.~M., {Stanke}, T., {et~al.} 2017, \aap, 600, A141

\bibitem[{{Koch} \& {Rosolowsky}(2015)}]{Koch15}
{Koch}, E.~W. \& {Rosolowsky}, E.~W. 2015, \mnras, 452, 3435

\bibitem[{{K{\"o}nyves} {et~al.}(2019){K{\"o}nyves}, {Andr{\'e}},
  {Arzoumanian}, {Schneider}, {Men'shchikov}, {Bontemps}, {Ladjelate},
  {Didelon}, {Pezzuto}, {Benedettini}, {Bracco}, {Di Francesco}, {Goodwin},
  {Rygl}, {Shimajiri}, {Spinoglio}, {Ward-Thompson}, \& {White}}]{Konyves19}
{K{\"o}nyves}, V., {Andr{\'e}}, P., {Arzoumanian}, D., {et~al.} 2019, \aap, in
  press (arXiv:1910.04053)

\bibitem[{{K{\"o}nyves} {et~al.}(2015){K{\"o}nyves}, {Andr{\'e}},
  {Men'shchikov}, {Palmeirim}, {Arzoumanian}, {Schneider}, {Roy}, {Didelon},
  {Maury}, {Shimajiri}, {Di Francesco}, {Bontemps}, {Peretto}, {Benedettini},
  {Bernard}, {Elia}, {Griffin}, {Hill}, {Kirk}, {Ladjelate}, {Marsh}, {Martin},
  {Motte}, {Nguy{\^e}n Luong}, {Pezzuto}, {Roussel}, {Rygl}, {Sadavoy},
  {Schisano}, {Spinoglio}, {Ward-Thompson}, \& {White}}]{Konyves15}
{K{\"o}nyves}, V., {Andr{\'e}}, P., {Men'shchikov}, A., {et~al.} 2015, \aap,
  584, A91

\bibitem[{{Lada} {et~al.}(2012){Lada}, {Forbrich}, {Lombardi}, \&
  {Alves}}]{Lada12}
{Lada}, C.~J., {Forbrich}, J., {Lombardi}, M., \& {Alves}, J.~F. 2012, \apj,
  745, 190

\bibitem[{{Lada} {et~al.}(2010){Lada}, {Lombardi}, \& {Alves}}]{Lada10}
{Lada}, C.~J., {Lombardi}, M., \& {Alves}, J.~F. 2010, \apj, 724, 687

\bibitem[{{Lee} {et~al.}(2017){Lee}, {Hennebelle}, \& {Chabrier}}]{Lee17}
{Lee}, Y.-N., {Hennebelle}, P., \& {Chabrier}, G. 2017, \apj, 847, 114

\bibitem[{{Longmore} {et~al.}(2013){Longmore}, {Bally}, {Testi}, {Purcell},
  {Walsh}, {Bressert}, {Pestalozzi}, {Molinari}, {Ott}, {Cortese}, {Battersby},
  {Murray}, {Lee}, {Kruijssen}, {Schisano}, \& {Elia}}]{Longmore13}
{Longmore}, S.~N., {Bally}, J., {Testi}, L., {et~al.} 2013, \mnras, 429, 987

\bibitem[{{Marsh} {et~al.}(2016){Marsh}, {Kirk}, {Andr{\'e}}, {Griffin},
  {K{\"o}nyves}, {Palmeirim}, {Men'shchikov}, {Ward-Thompson}, {Benedettini},
  {Bresnahan}, {di Francesco}, {Elia}, {Motte}, {Peretto}, {Pezzuto}, {Roy},
  {Sadavoy}, {Schneider}, {Spinoglio}, \& {White}}]{Marsh16}
{Marsh}, K.~A., {Kirk}, J.~M., {Andr{\'e}}, P., {et~al.} 2016, \mnras, 459, 342

\bibitem[{{Men'shchikov}(2013)}]{Menshchikov13}
{Men'shchikov}, A. 2013, \aap, 560, A63

\bibitem[{{Men'shchikov}(2017)}]{Menshchikov17}
{Men'shchikov}, A. 2017, \aap, 607, A64

\bibitem[{{Men'shchikov} {et~al.}(2012){Men'shchikov}, {Andr{\'e}}, {Didelon},
  {Motte}, {Hennemann}, \& {Schneider}}]{Menshchikov12}
{Men'shchikov}, A., {Andr{\'e}}, P., {Didelon}, P., {et~al.} 2012, \aap, 542,
  A81

\bibitem[{{Molinari} {et~al.}(2010){Molinari}, {Swinyard}, {Bally}, {Barlow},
  {Bernard}, {Martin}, {Moore}, {Noriega-Crespo}, {Plume}, {Testi}, {Zavagno},
  {Abergel}, {Ali}, {Anderson}, {Andr{\'e}}, {Baluteau}, {Battersby},
  {Beltr{\'a}n}, {Benedettini}, {Billot}, {Blommaert}, {Bontemps}, {Boulanger},
  {Brand}, {Brunt}, {Burton}, {Calzoletti}, {Carey}, {Caselli}, {Cesaroni},
  {Cernicharo}, {Chakrabarti}, {Chrysostomou}, {Cohen}, {Compiegne}, {de
  Bernardis}, {de Gasperis}, {di Giorgio}, {Elia}, {Faustini}, {Flagey},
  {Fukui}, {Fuller}, {Ganga}, {Garcia-Lario}, {Glenn}, {Goldsmith}, {Griffin},
  {Hoare}, {Huang}, {Ikhenaode}, {Joblin}, {Joncas}, {Juvela}, {Kirk},
  {Lagache}, {Li}, {Lim}, {Lord}, {Marengo}, {Marshall}, {Masi}, {Massi},
  {Matsuura}, {Minier}, {Miville-Desch{\^e}nes}, {Montier}, {Morgan}, {Motte},
  {Mottram}, {M{\"u}ller}, {Natoli}, {Neves}, {Olmi}, {Paladini}, {Paradis},
  {Parsons}, {Peretto}, {Pestalozzi}, {Pezzuto}, {Piacentini}, {Piazzo},
  {Polychroni}, {Pomar{\`e}s}, {Popescu}, {Reach}, {Ristorcelli}, {Robitaille},
  {Robitaille}, {Rod{\'o}n}, {Roy}, {Royer}, {Russeil}, {Saraceno}, {Sauvage},
  {Schilke}, {Schisano}, {Schneider}, {Schuller}, {Schulz}, {Sibthorpe},
  {Smith}, {Smith}, {Spinoglio}, {Stamatellos}, {Strafella}, {Stringfellow},
  {Sturm}, {Taylor}, {Thompson}, {Traficante}, {Tuffs}, {Umana}, {Valenziano},
  {Vavrek}, {Veneziani}, {Viti}, {Waelkens}, {Ward-Thompson}, {White},
  {Wilcock}, {Wyrowski}, {Yorke}, \& {Zhang}}]{Molinari10}
{Molinari}, S., {Swinyard}, B., {Bally}, J., {et~al.} 2010, \aap, 518, L100

\bibitem[{{Myers}(2009)}]{Myers09}
{Myers}, P.~C. 2009, \apj, 700, 1609

\bibitem[{{Ohashi} {et~al.}(2016){Ohashi}, {Sanhueza}, {Chen}, {Zhang},
  {Busquet}, {Nakamura}, {Palau}, \& {Tatematsu}}]{Ohashi16}
{Ohashi}, S., {Sanhueza}, P., {Chen}, H.-R.~V., {et~al.} 2016, \apj, 833, 209

\bibitem[{{Palmeirim} {et~al.}(2013){Palmeirim}, {Andr{\'e}}, {Kirk},
  {Ward-Thompson}, {Arzoumanian}, {K{\"o}nyves}, {Didelon}, {Schneider},
  {Benedettini}, {Bontemps}, {Di Francesco}, {Elia}, {Griffin}, {Hennemann},
  {Hill}, {Martin}, {Men'shchikov}, {Molinari}, {Motte}, {Nguyen Luong},
  {Nutter}, {Peretto}, {Pezzuto}, {Roy}, {Rygl}, {Spinoglio}, \&
  {White}}]{Palmeirim13}
{Palmeirim}, P., {Andr{\'e}}, P., {Kirk}, J., {et~al.} 2013, \aap, 550, A38

\bibitem[{{Peretto} {et~al.}(2013){Peretto}, {Fuller}, {Duarte-Cabral},
  {Avison}, {Hennebelle}, {Pineda}, {Andr{\'e}}, {Bontemps}, {Motte},
  {Schneider}, \& {Molinari}}]{Peretto13}
{Peretto}, N., {Fuller}, G.~A., {Duarte-Cabral}, A., {et~al.} 2013, \aap, 555,
  A112

\bibitem[{{Persi} \& {Tapia}(2008)}]{Persi08}
{Persi}, P. \& {Tapia}, M. 2008, {Star Formation in NGC 6334}, ed.
  B.~{Reipurth}, 456

\bibitem[{{Rodriguez} {et~al.}(1982){Rodriguez}, {Canto}, \&
  {Moran}}]{Rodriguez82}
{Rodriguez}, L.~F., {Canto}, J., \& {Moran}, J.~M. 1982, \apj, 255, 103

\bibitem[{{Roy} {et~al.}(2014){Roy}, {Andr{\'e}}, {Palmeirim}, {Attard},
  {K{\"o}nyves}, {Schneider}, {Peretto}, {Men'shchikov}, {Ward-Thompson},
  {Kirk}, {Griffin}, {Marsh}, {Abergel}, {Arzoumanian}, {Benedettini}, {Hill},
  {Motte}, {Nguyen Luong}, {Pezzuto}, {Rivera-Ingraham}, {Roussel}, {Rygl},
  {Spinoglio}, {Stamatellos}, \& {White}}]{Roy14}
{Roy}, A., {Andr{\'e}}, P., {Palmeirim}, P., {et~al.} 2014, \aap, 562, A138

\bibitem[{{Russeil} {et~al.}(2013){Russeil}, {Schneider}, {Anderson},
  {Zavagno}, {Molinari}, {Persi}, {Bontemps}, {Motte}, {Ossenkopf},
  {Andr{\'e}}, {Arzoumanian}, {Bernard}, {Deharveng}, {Didelon}, {Di
  Francesco}, {Elia}, {Hennemann}, {Hill}, {K{\"o}nyves}, {Li}, {Martin},
  {Nguyen Luong}, {Peretto}, {Pezzuto}, {Polychroni}, {Roussel}, {Rygl},
  {Spinoglio}, {Testi}, {Tig{\'e}}, {Vavrek}, {Ward-Thompson}, \&
  {White}}]{Russeil13}
{Russeil}, D., {Schneider}, N., {Anderson}, L.~D., {et~al.} 2013, \aap, 554,
  A42

\bibitem[{{Schisano} {et~al.}(2014){Schisano}, {Rygl}, {Molinari}, {Busquet},
  {Elia}, {Pestalozzi}, {Polychroni}, {Billot}, {Carey}, {Paladini},
  {Noriega-Crespo}, {Moore}, {Plume}, {Glover}, \&
  {V{\'a}zquez-Semadeni}}]{Schisano14}
{Schisano}, E., {Rygl}, K.~L.~J., {Molinari}, S., {et~al.} 2014, \apj, 791, 27

\bibitem[{{Shimajiri} {et~al.}(2017){Shimajiri}, {Andr{\'e}}, {Braine},
  {K{\"o}nyves}, {Schneider}, {Bontemps}, {Ladjelate}, {Roy}, {Gao}, \&
  {Chen}}]{Shimajiri17}
{Shimajiri}, Y., {Andr{\'e}}, P., {Braine}, J., {et~al.} 2017, \aap, 604, A74

\bibitem[{{Shimajiri} {et~al.}(2019){Shimajiri}, {Andr{\'e}}, {Palmeirim},
  {Arzoumanian}, {Bracco}, {K{\"o}nyves}, {Ntormousi}, \&
  {Ladjelate}}]{Shimajiri18}
{Shimajiri}, Y., {Andr{\'e}}, P., {Palmeirim}, P., {et~al.} 2019, \aap, 623,
  A16

\bibitem[{{Shirley}(2015)}]{Shirley15}
{Shirley}, Y.~L. 2015, \pasp, 127, 299

\bibitem[{{Smith} {et~al.}(2014){Smith}, {Glover}, \& {Klessen}}]{Smith14}
{Smith}, R.~J., {Glover}, S.~C.~O., \& {Klessen}, R.~S. 2014, \mnras, 445, 2900

\bibitem[{{Stutzki} \& {Guesten}(1990)}]{Stutzki90}
{Stutzki}, J. \& {Guesten}, R. 1990, \apj, 356, 513

\bibitem[{{Tafalla} \& {Hacar}(2015)}]{Tafalla15}
{Tafalla}, M. \& {Hacar}, A. 2015, \aap, 574, A104

\bibitem[{{Takahashi} {et~al.}(2013){Takahashi}, {Ho}, {Teixeira}, {Zapata}, \&
  {Su}}]{Takahashi13}
{Takahashi}, S., {Ho}, P.~T.~P., {Teixeira}, P.~S., {Zapata}, L.~A., \& {Su},
  Y.-N. 2013, \apj, 763, 57

\bibitem[{{Teixeira} {et~al.}(2016){Teixeira}, {Takahashi}, {Zapata}, \&
  {Ho}}]{Teixeira16}
{Teixeira}, P.~S., {Takahashi}, S., {Zapata}, L.~A., \& {Ho}, P.~T.~P. 2016,
  \aap, 587, A47

\bibitem[{{Tig{\'e}} {et~al.}(2017){Tig{\'e}}, {Motte}, {Russeil}, {Zavagno},
  {Hennemann}, {Schneider}, {Hill}, {Nguyen Luong}, {Di Francesco}, {Bontemps},
  {Louvet}, {Didelon}, {K{\"o}nyves}, {Andr{\'e}}, {Leuleu}, {Bardagi},
  {Anderson}, {Arzoumanian}, {Benedettini}, {Bernard}, {Elia}, {Figueira},
  {Kirk}, {Martin}, {Minier}, {Molinari}, {Nony}, {Persi}, {Pezzuto},
  {Polychroni}, {Rayner}, {Rivera-Ingraham}, {Roussel}, {Rygl}, {Spinoglio}, \&
  {White}}]{Tige17}
{Tig{\'e}}, J., {Motte}, F., {Russeil}, D., {et~al.} 2017, \aap, 602, A77

\bibitem[{{V{\'a}zquez-Semadeni} {et~al.}(2019){V{\'a}zquez-Semadeni}, {Palau},
  {Ballesteros-Paredes}, {G{\'o}mez}, \& {Zamora-Avil{\'e}s}}]{Vazquez19}
{V{\'a}zquez-Semadeni}, E., {Palau}, A., {Ballesteros-Paredes}, J.,
  {G{\'o}mez}, G.~C., \& {Zamora-Avil{\'e}s}, M. 2019, \mnras, submitted
  (astro-ph/1903.11247)

\bibitem[{{Willis} {et~al.}(2013){Willis}, {Marengo}, {Allen}, {Fazio},
  {Smith}, \& {Carey}}]{Willis13}
{Willis}, S., {Marengo}, M., {Allen}, L., {et~al.} 2013, \apj, 778, 96

\bibitem[{{Wilner} \& {Welch}(1994)}]{Wilner94}
{Wilner}, D.~J. \& {Welch}, W.~J. 1994, \apj, 427, 898

\bibitem[{{Zhang} {et~al.}(2019){Zhang}, {Kainulainen}, {Mattern}, {Fang}, \&
  {Henning}}]{Zhang19}
{Zhang}, M., {Kainulainen}, J., {Mattern}, M., {Fang}, M., \& {Henning}, T.
  2019, \aap, 622, A52

\bibitem[{{Zhang} {et~al.}(2015){Zhang}, {Wang}, {Lu}, \&
  {Jim{\'e}nez-Serra}}]{Zhang15}
{Zhang}, Q., {Wang}, K., {Lu}, X., \& {Jim{\'e}nez-Serra}, I. 2015, \apj, 804,
  141

\end{thebibliography}

\appendix
\section{Complementary figures}\label{Sect:Complementary}

\begin{figure*}
\centering
\includegraphics[width=180mm, angle=0]{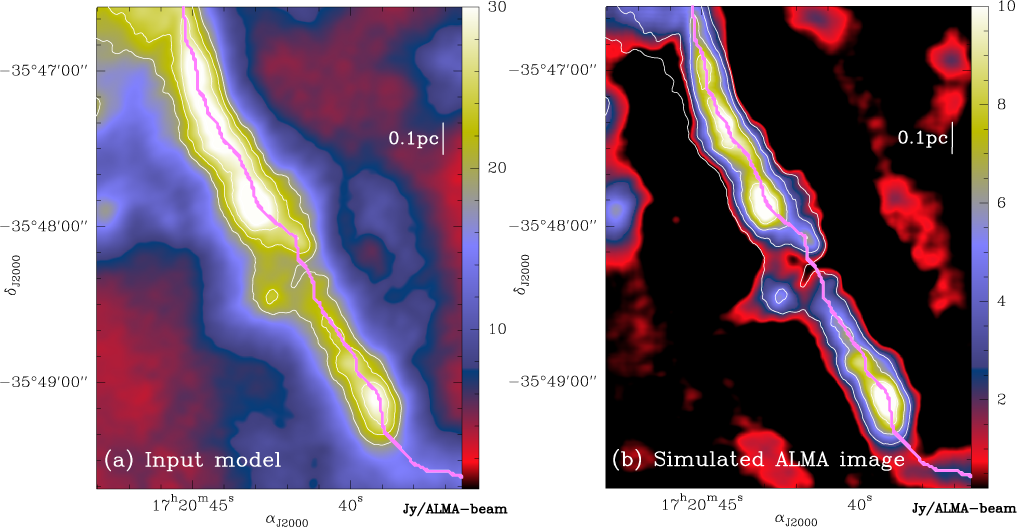}
\caption{Simulations of ALMA$+$ACA observations performed with CASA and using the ArT\'eMiS$+$SPIRE 350\,$\mu$m image \citep{Andre16} as input model.
(a) Input model: ArT\'eMiS 350\,$\mu$m map in units of Jy/ALMA-beam. (b) Simulated ALMA$+$ACA image using the same set of baselines (i.e., u--v coverage) as the real data. 
While extended ($> 37\arcsec $) emission is filtered out, we note that interferometric-filtering effects do not generate spurious sub-structures within the filament. 
}
\label{fig_alma_sim}
\end{figure*}

\begin{figure*}
\centering
\includegraphics[width=180mm, angle=0]{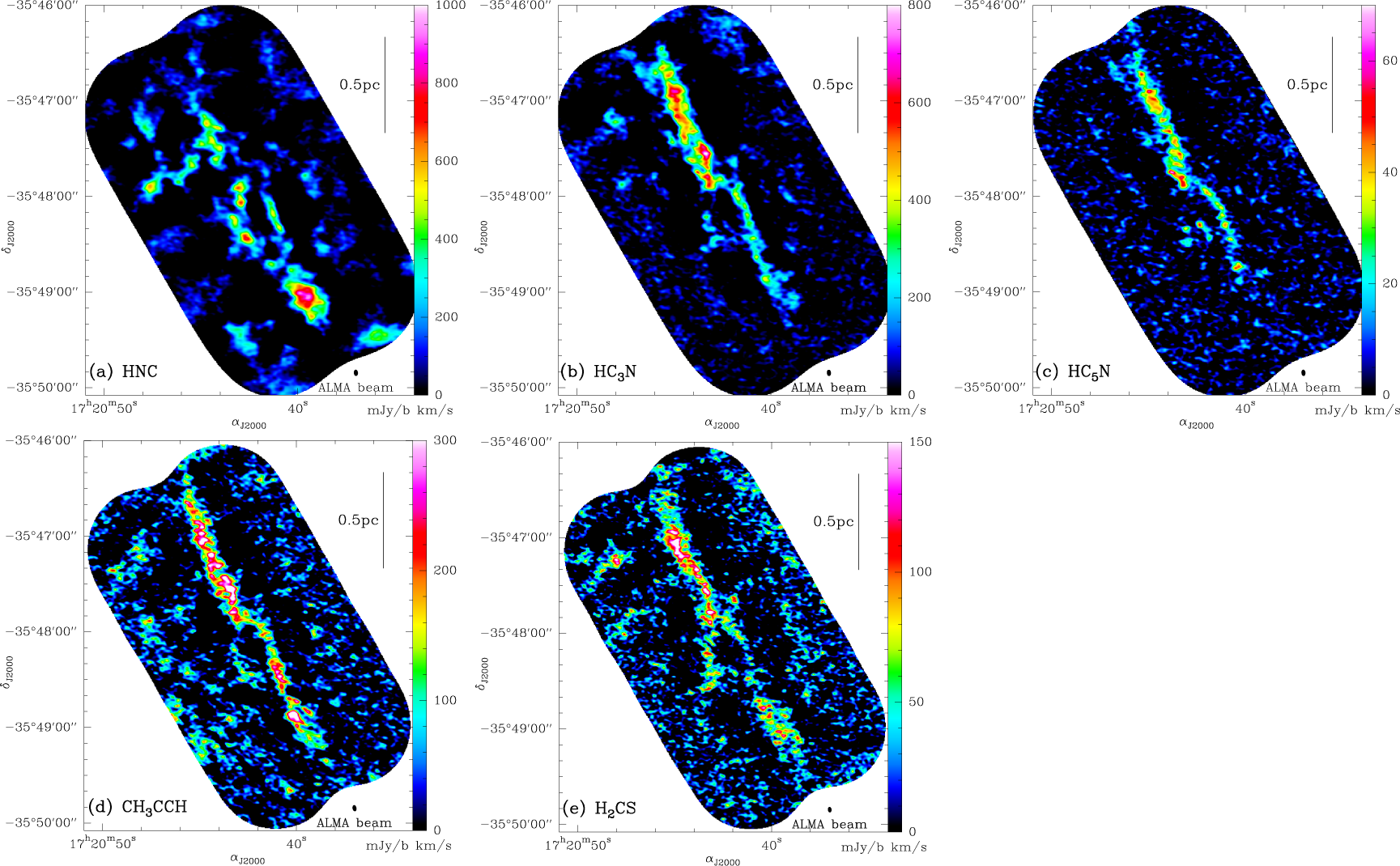}
\caption{Integrated intensity maps of the NGC~6334 main filament region in HNC(1--0) ($a$),  HC$_3$N(10--9) ($b$), HC$_5$N(36--35) ($c$) , CH$_3$CCH(6$_0$--5$_0$) ($d$), 
and H$_2$CS(3$_{1,2}$--2$_{1,1}$) ($e$) with ALMA 12m+7m.
}
\label{figs_otherlines}
\end{figure*}

Figure \ref{fig_alma_sim} and Figure~\ref{figs_otherlines} are complementary figures.

\section{Treatment of multiple velocity components in the N$_2$H$^+$(1--0) spectra}\label{Sect:Contamination}

In order to disentangle between the N$_2$H$^+$(1--0) HFS 1 emission from the main velocity component at $V_{\rm LSR} = -6\,$km s$^{-1}$
and the HFS 2-7 emission from other velocity components, we compared the observations with 
synthetic N$_2$H$^+$ emission models including two velocity components along the line of the sight. 

Assuming the same excitation temperature and the same linewidth for all hyperfine transitions (HFS 1-7),
and Gaussian line opacity profiles as a function of velocity, 
the N$_2$H$^+$ temperature and the N$_2$H$^+$ opacity at velocity $V$ are given by:

\begin{equation}
T(V) = \frac{p_1}{p_4}\Bigl(1-e^{-\tau(V)}\Bigr),
\end{equation}

\noindent where

\begin{equation}
\tau(V) = p_4 \times \Sigma_{i=1}^{7} r_i \exp \Bigl[-4\ln 2 \Bigl(\frac{V-V_i-p_2}{p_3}\Bigr)^2 \Bigr]
\end{equation}

\noindent where the parameters $p_1$, $p_2$, $p_3$, $p_4$ represent $T$ $\times$ $\tau$, $V_{\rm LSR}$, linewidth, and opacity $\tau$ 
for the main HFS component (HFS 3)\footnote{See \url{https://www.iram.fr/IRAMFR/GILDAS/doc/pdf/class.pdf} for details.}.

\begin{figure*}
\centering
\includegraphics[width=60mm, angle=0]{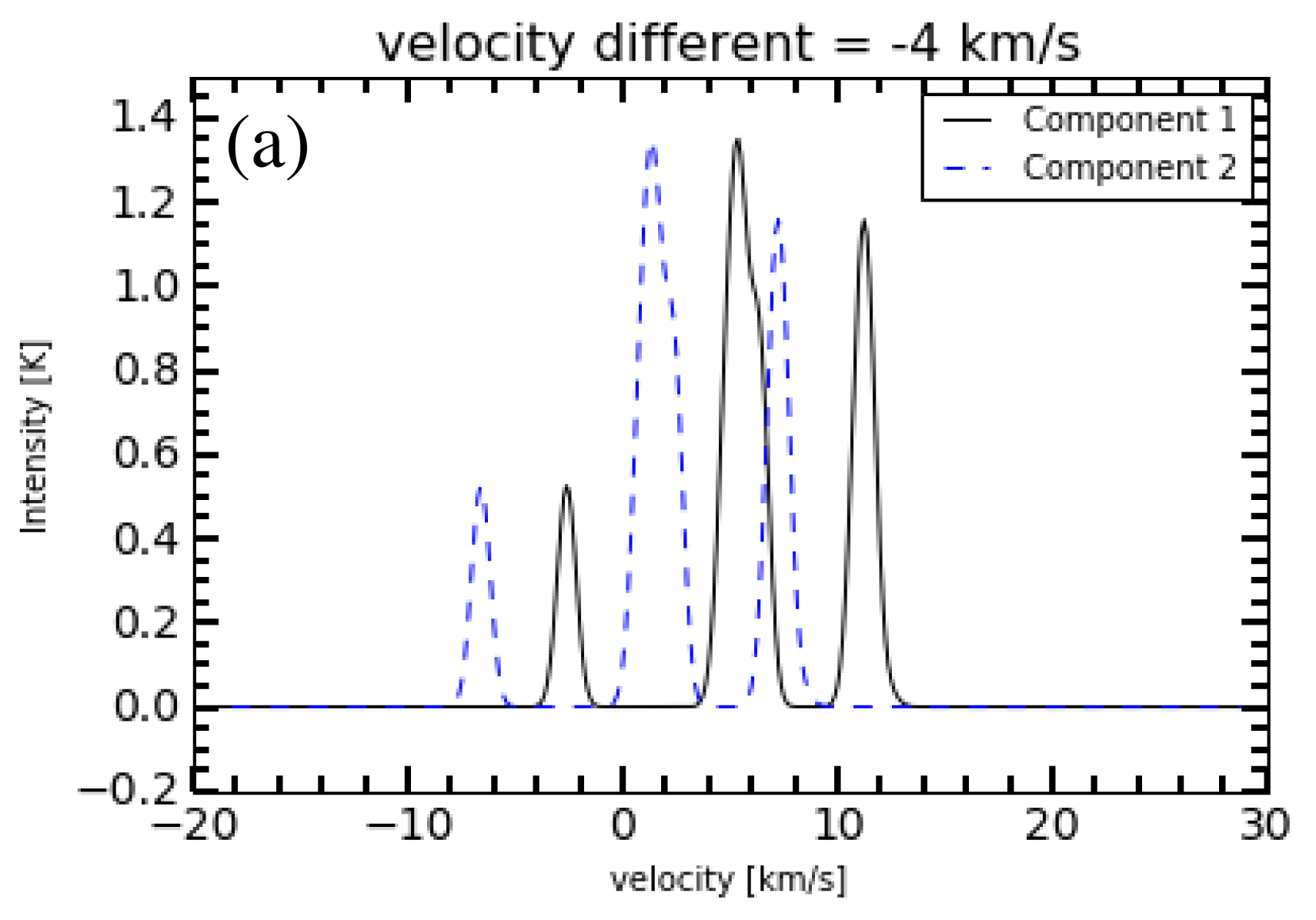}
\includegraphics[width=60mm, angle=0]{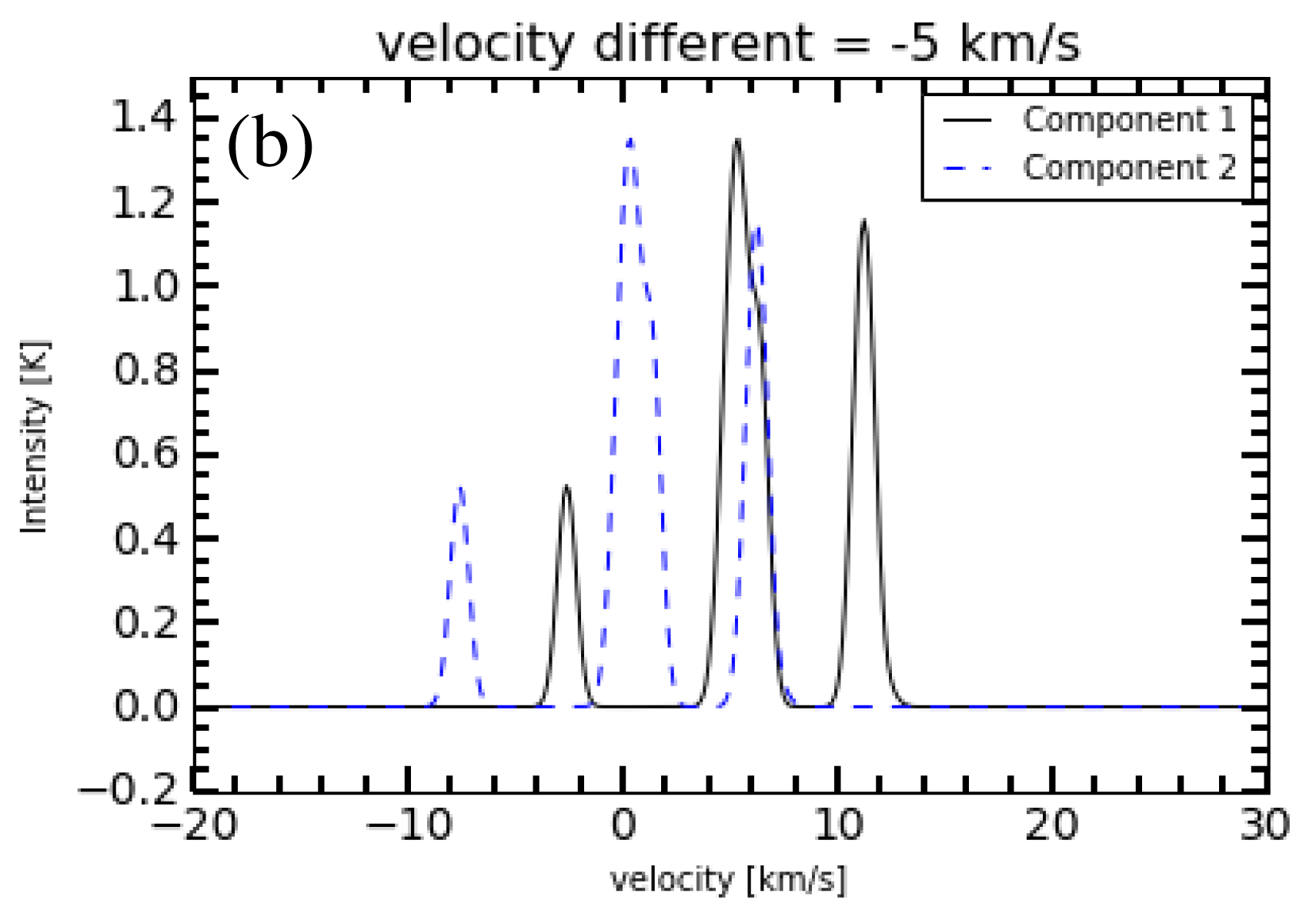}
\includegraphics[width=60mm, angle=0]{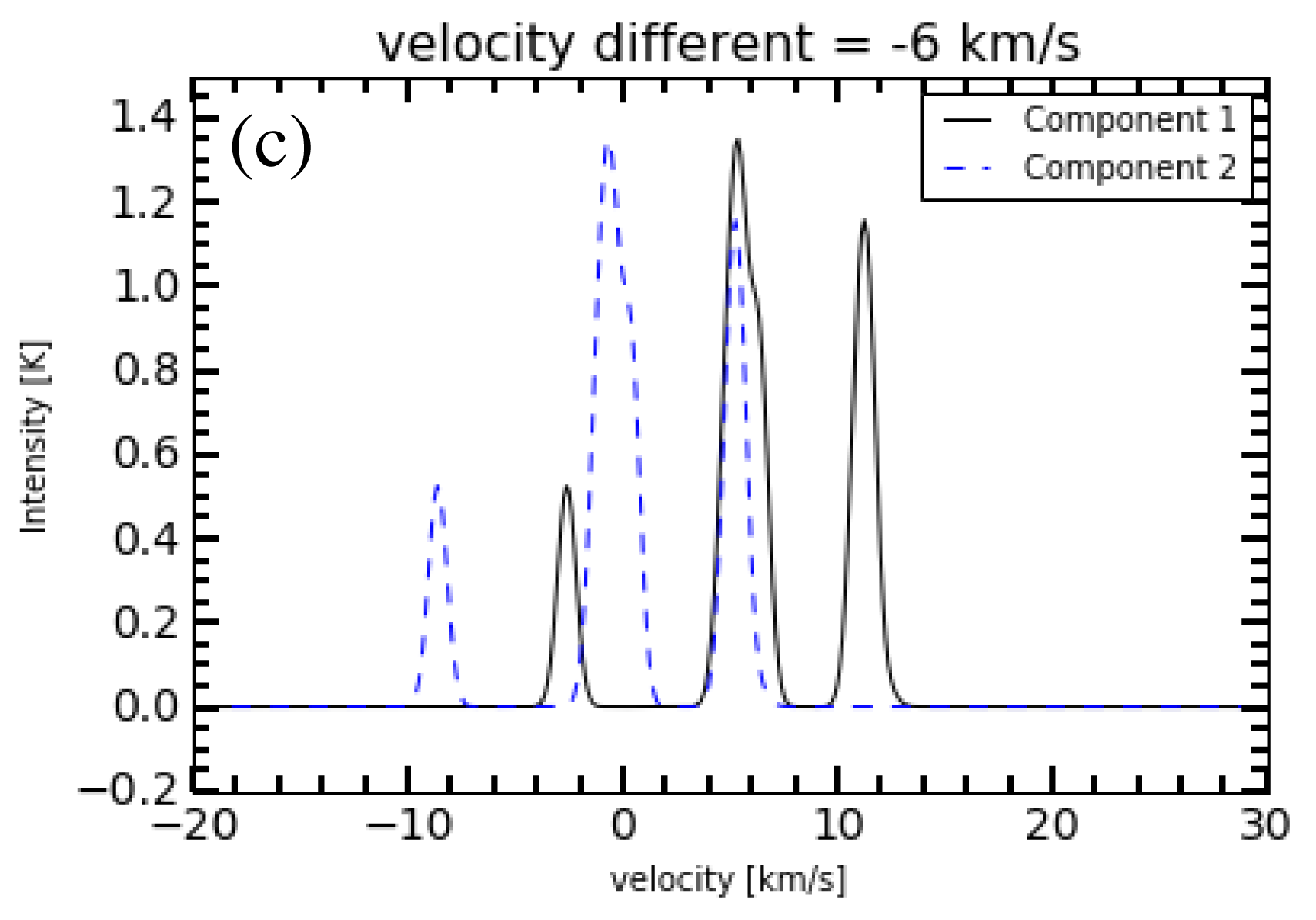}
\caption{Schematic pictures of the contamination of the main velocity component the N$_{2}$H$^+$ (1--0) emission (1, solid line) 
by the HFS 2-7 emission from  other velocity components. 
The velocity differences between component 1 and component 2 are 4, 5, and 6 km s$^{-1}$
in panels (${\it a}$), (${\it b}$), and (${\it c}$), respectively. 
}
\label{figs:N2Hp_contamination}
\end{figure*}

\begin{figure*}
\centering
\includegraphics[width=190mm, angle=0]{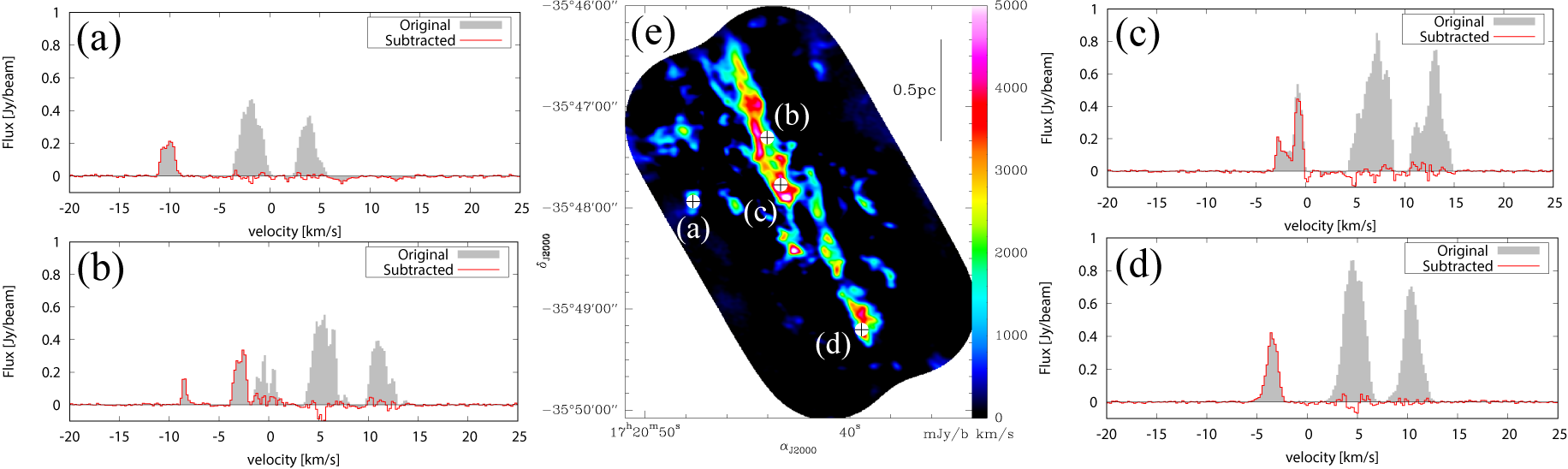}
\caption{Examples of N$_2$H$^+$(1--0) spectra before (gray)
and after (red) subtraction of contaminating emission from hyperfine components HFS~2--7 (see Sect. \ref{sect:subtraction_uncomp}). 
In panel ($e$), black plus symbols within white circles mark the positions of the spectra shown in the other panels.
}
\label{figs:n2h+_spectrum_removal}
\end{figure*}

Figure \ref{figs:N2Hp_contamination} shows model spectra in the case of two velocity components (components 1 and 2) along the line of the sight. 
To make it simple, we assumed $p_1$=1.0, $p_3$=1.0 km s$^{-1}$, and $p_4$=0.2 for both velocity components. 
We adopted a LSR velocity $p_2$=$-$2.6 km s$^{-1}$ for component 1. 
For component 2, we added a velocity offset of $-$4, $-$5, and $-$6 km s$^{-1}$ to $p_2$=$-$2.6 km s$^{-1}$.  
It can be seen that, for an offset of $-$4 km s$^{-1}$, the HFS 1 emission of component 1 is not contaminated by the HFS 2-7 emission of component 2. 
However, for offsets between $-$5 and $-$6 km s$^{-1}$, the HFS 1 emission of component 1 can be significantly contaminated. 
Thus, to avoid contamination by the HFS 2-7 emission from other velocity components, 
we performed a multi-component HFS fitting of the observed spectra for all pixels 
showing evidence of more than one velocity component along the line of sight
with blueshifted velocity offsets $>$ 4 km s$^{-1}$ in magnitude. 
This HFS fitting with $N$ HFS components was applied to each pixel, where $N$=1, 2, or 3. 
A initial HFS fitting step was first performed with $N$=1. Whenever the peak signal to noise ratio (S/N) of the residual intensity 
was lower than 10, the fit was deemed to be acceptable and the corresponding spectrum was assumed to consist 
of $N=1$ Gaussian components. Whenever the peak S/N of the residuals was larger than 10, 
an additional HFS fitting step was performed with $N$+1 components. 
In practice, all pixels with detected N$_2$H$^+$ emission were fit with $N$ $\leq$ 3 components.
The fit HFS 2-7 emission was then subtracted from the observed spectra. 
Accordingly, the resulting data cube only has  significant emission from HFS 1 as shown in Fig.~\ref{figs:n2h+_spectrum_removal}.

\section{Post-extraction selection of \textsl{getsources} results}\label{Sect:getsources_post_selection_detail}

In Sect. \ref{sect:getsources}, we applied the \textsl{getsources} algorithm to extract compact sources from the ALMA 3.1 mm continuum map 
and N$_2$H$^+$ data cube. 
Here, we describe the additional criteria used to select robust compact sources from the \textsl{getsources} extraction results. 

The \textsl{getsources} code identified a total of 49 compact sources in  the 3.1 mm continuum map. 
First, we removed all sources lying closer than 3 ALMA beams from the edge of the map. 
With this criterion, 3 of the 49 initial sources were removed.  
We also removed 6 sources identified by \textsl{getsources} with a peak intensity below $<$ 5$\sigma$. 
As a result of these two selection steps, only 40 robust continuum sources remained. 

Then, in order to investigate the potential association of 
these compact 3.1~mm continuum sources with compact N$_2$H$^+$ emission/sources, 
we also applied the \textsl{getsources} algorithm to each velocity channel map in the N$_2$H$^+$ data cube. 
In this case, we removed extracted sources with a peak intensity below $<$3$\sigma$ in individual channel maps.
We selected only compact N$_2$H$^+$ sources detected in at least two contiguous velocity channels (where the channel width corresponds to 0.2 km s$^{-1}$)
and lying within one ALMA beam ($\sim$$3\arcsec$ or $\sim$0.02 pc) 
of a 3.1~mm continuum source.
As a result, we found that 21 (or 81\%) of the 26 continuum cores are associated with 
compact N$_2$H$^+$ emission.

\section{Post-extraction selection of \textsl{getfilaments} results}\label{Sect:getfilament_post_selection_detail}

In Sect. \ref{sect:Post-selection}, we summarized the additional criteria used to select robust filamentary (sub-)structures from the \textsl{getfilaments} results. 
Here, we provide more details. 

\begin{figure*}[htp]
\centering
\includegraphics[width= 160mm, angle=0]{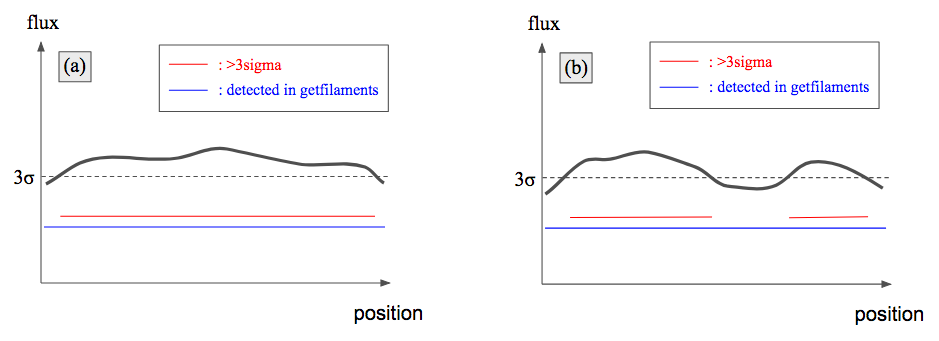}
\caption{Schematic illustration of the first criterion used in the post-selection of \textsl{getfilaments} results. 
In panel (a), $>$90\% of the pixels on the crest of the filamentary structure 
are detected above 3$\sigma$. In panel (b), $<$90\% of the pixels on the crest 
of the filamentary structure identified with \textsl{getfilaments} 
are detected above 3$\sigma$.  }
\label{figs:post_selection_1st}
\end{figure*}

Our first criterion was to impose 
that $>$90\% of the pixels on the crest of a robust filamentary structure 
are detected above the 3$\sigma$ level. 
In Fig. \ref{figs:post_selection_1st} (a),  for instance, $>$90\% of the pixels on the crest of the filamentary structure 
are detected above 3$\sigma$. Thus, this filamentary structure would pass the first criterion. 
In Fig. \ref{figs:post_selection_1st} (b), $<$90\% of the pixels on the crest of the filamentary structure 
are detected above 3$\sigma$. 
Thus, this structure would be regarded as a fake filamentary structure and would be removed from the detection list. 

Then, we connected filamentary structures detected in distinct velocity channels by comparing 
their spatial distributions in adjacent velocity-channel maps:
Whenever two structures detected in adjacent velocity-channel maps
overlapped over $>$60\% of the pixels, they were connected together.

\begin{figure*}
\centering
\includegraphics[width= 170mm, angle=0]{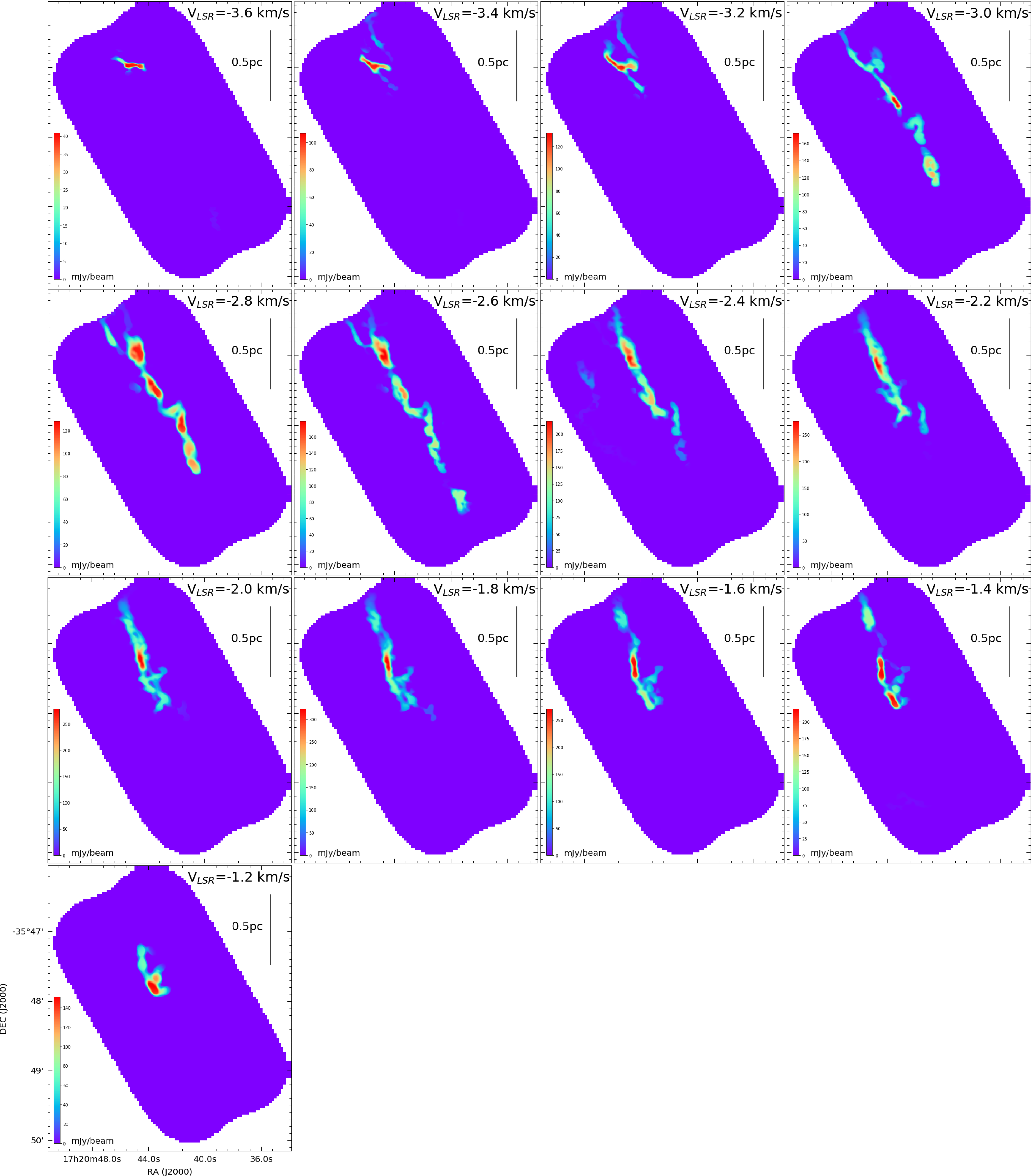}
\caption{N$_2$H$^+$ velocity channel maps after subtraction of compact N$_2$H$^+$ emission 
using the \textsl{getfilaments} algorithm.
 }
\label{figs:channel_map_subtracted}
\end{figure*}

\begin{figure*}
\centering
\includegraphics[width= 170mm, angle=0]{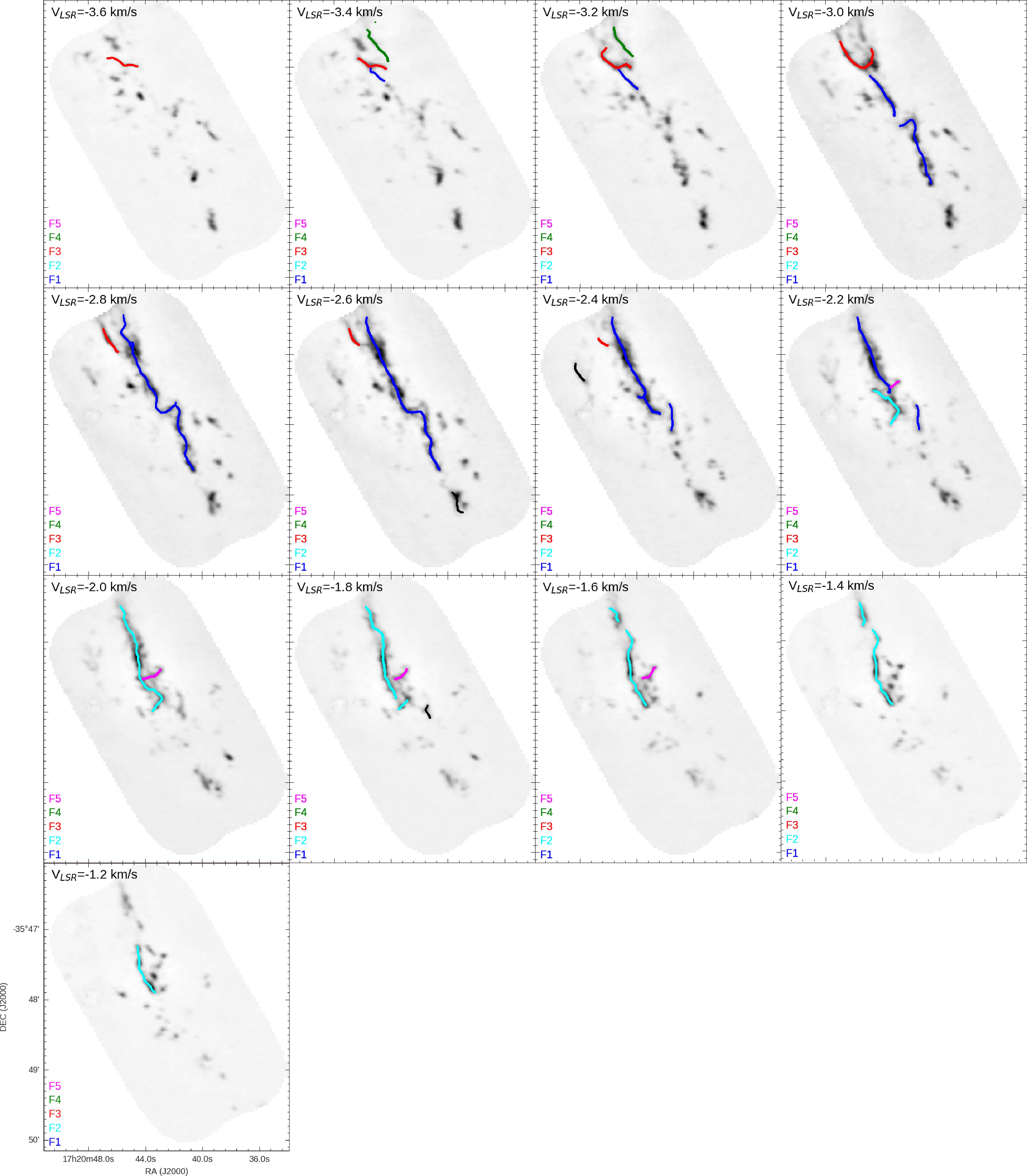}
\caption{Crests of the filamentary structures identified in the N$_2$H$^+$ velocity channel maps. 
In each panel, the grayscale shows the distribution of N$_2$H$^+$ emission. 
Dark blue, light blue, red, green, and magenta curves mark the crests of F-1, F-2, F-3, F-4, F-5, respectively. 
Black curves indicate  removed filamentary structures which are detected in only one velocity channel.  }
\label{figs:channel_map_crest}
\end{figure*}

In practice, our step-by-step procedure to associate velocity-coherent features seen in adjacent velocity channels,  
from the most blueshifted to the most redshifted channel, can be described  as follows. 
Figures \ref{figs:channel_map_subtracted} and \ref{figs:channel_map_crest} show the velocity channel maps of the 
compact-source-subtracted N$_{2}$H$^+$ emission cube that were used for the identification of filamentary structures 
with \textsl{getfilaments} and the crests of these structures 
in each velocity channel (after post-extraction selection).\\
In the channel map at $V_{\rm LSR}$ = $-$3.6 km s$^{-1}$, 
one filamentary structure is detected. This filamentary structure is labeled 
F-3\footnote{In Table~\ref{table:N$_2$H$^+$_filament}, the identified fiber-like features are listed by decreasing order of length.}.\\ 
In the channel maps at $V_{\rm LSR}$ = $-$3.4 km s$^{-1}$ and $-$3.2 km s$^{-1}$, three filamentary structures are detected in both channels.  
The spatial distributions of the detected filamentary structures are consistent in both channels. 
The filamentary structure whose distribution coincides with that of F-3 at $-$3.6 km s$^{-1}$ is identified as F-3. 
The other two filamentary structures are labeled F-4 and F-1.\\

In the channel map at $V_{\rm LSR}$ = $-$3.0 km s$^{-1}$, three discrete filaments are detected. 
The spatial distribution of one part of the northernmost filamentary structure coincides with the distribution of F-3 at $-$3.6 to $-$3.2 km s$^{-1}$. 
Thus, this filamentary structure is identified as being F-3. 
The spatial distribution of another one overlaps that of F-1  at $-$3.4 km s$^{-1}$.
The spatial distribution of the third structure coincides with the distribution of one of the filamentary structures detected in the next velocity channel ($-$2.8 km s$^{-1}$). 
Thus, these features are identified as being the same filamentary structure (F-3).\\

In the channel maps for $V_{\rm LSR}$ = $-$2.8 and $-$2.6 km s$^{-1}$, 
two filamentary structures corresponding to F-3 and F-1  are detected. 
At $-$2.6 km s$^{-1}$, the filamentary structure shown as a black curve was removed as it is only detected in this channel.\\

In the channel map at $V_{\rm LSR}$ = $-$2.4 km s$^{-1}$, 
three filamentary structures are detected. One of them was removed since it is only detected in this channel. 
The other two discrete structures are identified as F-1 since their distributions 
overlap with that of F-1 at $-$2.6 km s$^{-1}$.\\

In the channel map at $V_{\rm LSR}$ = $-$2.2 km s$^{-1}$, 
four components can be  identified. Two discrete structures correspond to F-1  since their distributions overlap with that of F-1  at $-$2.4 km s$^{-1}$.
The filamentary structure labeled as F-3 is roughly perpendicular to the other filamentary structures. 
The spatial distribution of the structure labeled F-2  partly overlaps with the distribution of F-1 at $-$2.4 km s$^{-1}$. 
Thus, there is a possibility that  F-1 and F-2  are actually part of the same physical structure. 
But the northern portion of F-2 lies slightly  to the west of F-1,  while the southern part of F-2 lies to the east of F-1. 
Moreover, at positions where F-1 and F-2 overlap in the plane of the sky, the N$_2$H$^+$ spectra clearly exhibit distinct velocity components  
[see, e.g., positions (c) and (d) in Fig.~\ref{figs:n2h+_fiber_spectrum}]. Therefore, F-1 and F-2 seem to be distinct velocity-coherent features.\\

In the channel maps at $V_{\rm LSR}$ = $-$2.0 to $-$1.6 km s$^{-1}$, 
two filamentary structures (F-5 and F-2) can be identified. 
But F-2  lies slightly to the west of F-1 in the northern part (DEC > -35:47:30), 
while the southern part of F-2  lies to the east of F-1  in the southern part (DEC < -35:47:30). 
At $-$1.8 km s$^{-1}$, one structure was removed since it is detected in only one channel.

\section{Origin of the highly blueshifted N$_2$H$^+$ blobs}\label{Sect:origin_high_blue}

We detected highly blue-shifted N$_2$H$^+$ blobs, with LSR velocities up to $-$12 km s$^{-1}$ compared to the filament systemic velocity of $-$2.6 km s$^{-1}$ (see Sect. \ref{Sect:velo_structure}). 
There are at least two possibilities for their origin.

The first possibility is that the highly blue-shifted N$_2$H$^+$ blobs are associated with cloud-cloud collision. 
\citet{Fukui18} found the presence of highly blue-shifted emission up to $-$20 km s$^{-1}$ on $\sim \,$10-pc scales in their wide-field NANTEN2 $^{12}$CO (1--0, 2--1) data around Galactic longitude of 
$l =\ $351.1--351.5~deg which includes our ALMA observed area. 
They suggested that a large-scale cloud-cloud collision occurred a few Myr ago and produced this highly blue-shifted $^{12}$CO emission in the NGC~6334 complex.

The second possibility is that the highly blue-shifted N$_2$H$^+$ emission blobs are associated with the expanding shell produced by source D.  
As described in Sect. \ref{result:cont}, the distribution of the 3.1~mm continuum emission is shaped like a shell around source D. 
Most of the highly blue-shifted N$_2$H$^+$ blobs are distributed on the outskirts of this shell-like structure. 
While the northern part of the NGC 6334 filament is seen in absorption in the {\it Spitzer} 8 $\mu$m 
map, such absorption is not seen in the southern part (around the shell-like structure). 
This suggests that the southern part of the NGC 6334 filament is in the back of the exciting star (source D). 
If the N$_2$H$^+$ blobs were associated with the shell-like structure, their velocity would be expected to be 
redshifted compared to the surrounding gas material.  This is not  consistent with the fact that the N$_2$H$^+$ blobs are blueshifted.

We conclude that the highly blueshifted N$_2$H$^+$ blobs are more likely related 
to the cloud-cloud collision scenario proposed by \citet{Fukui18}. 

\begin{figure}
\centering
\includegraphics[width=90mm, angle=0]{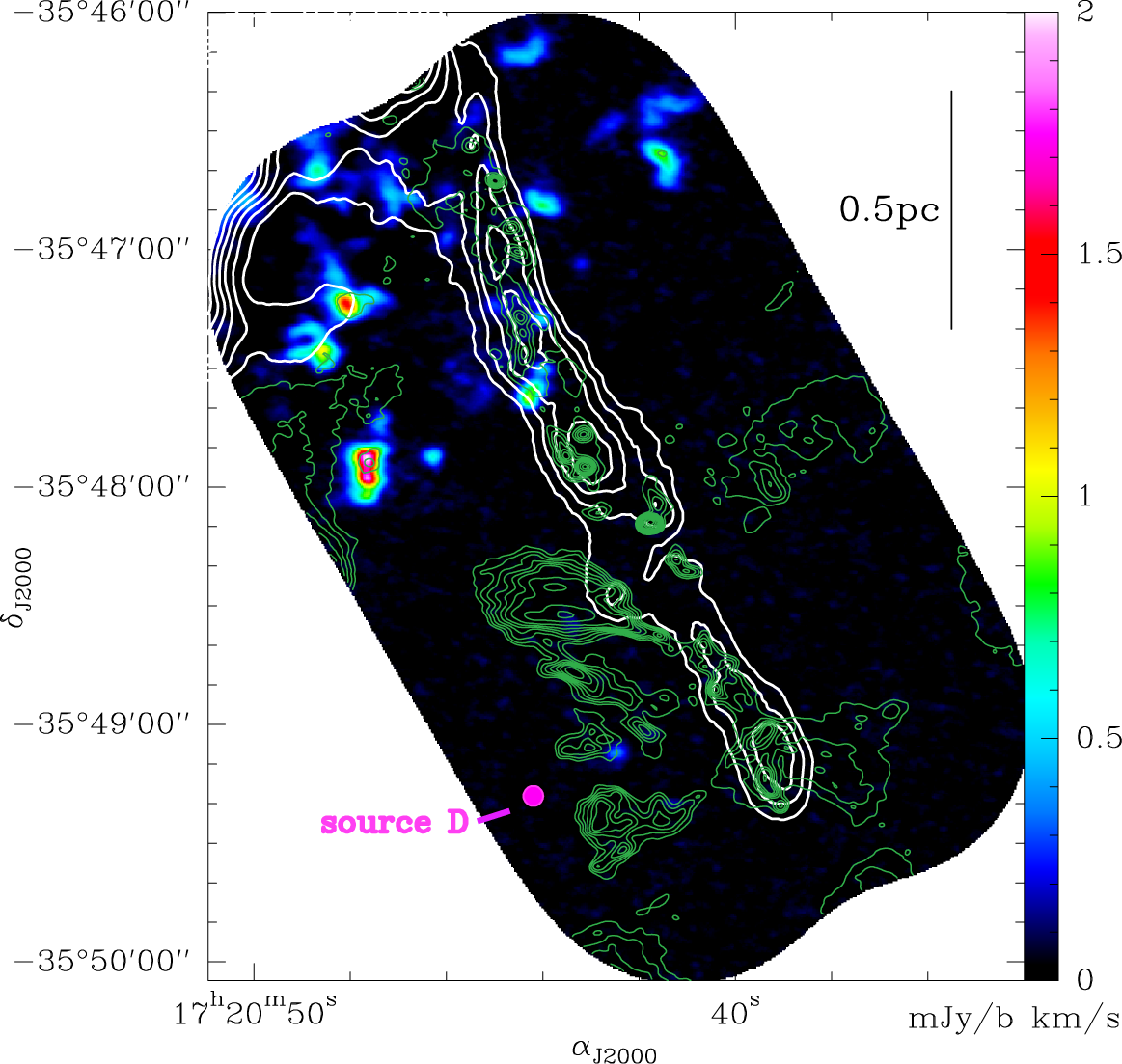}
\caption{ALMA 12m+7m map of the N$_2$H$^+$(1--0) highly blue shifted emission integrated from $-$12 km s$^{-1}$ to $-$6 km s$^{-1}$ (color scale).   
The superimposed white contours correspond to ArT$\acute{\rm e}$MiS 350 $\mu$m continuum levels 
of 10, 12, 14, 16, 18, and 20 Jy/8\arcsec -beam. 
Green contours of ALMA 3.1 mm continuum emission are also overlaid with levels of 2, 4, 6, 8, 10, 15, 20, 25, 30, 35, and 40$\sigma$, where 1$\sigma$ = 0.2 mJy/ALMA-beam.
The magenta filled circle marks the position of radio continuum source D (\citealp{Rodriguez82}) associated with a compact HII region.}
\label{figs:n2h+_blue}
\end{figure}

\section{Comparison between GAUSSCLUMPS, \textsl{getsources}, and REINOHLD extractions of clumps in the ArT\'eMiS map}\label{Sect:comparison_IDs}

\begin{figure}
\centering
\includegraphics[width=90mm, angle=0]{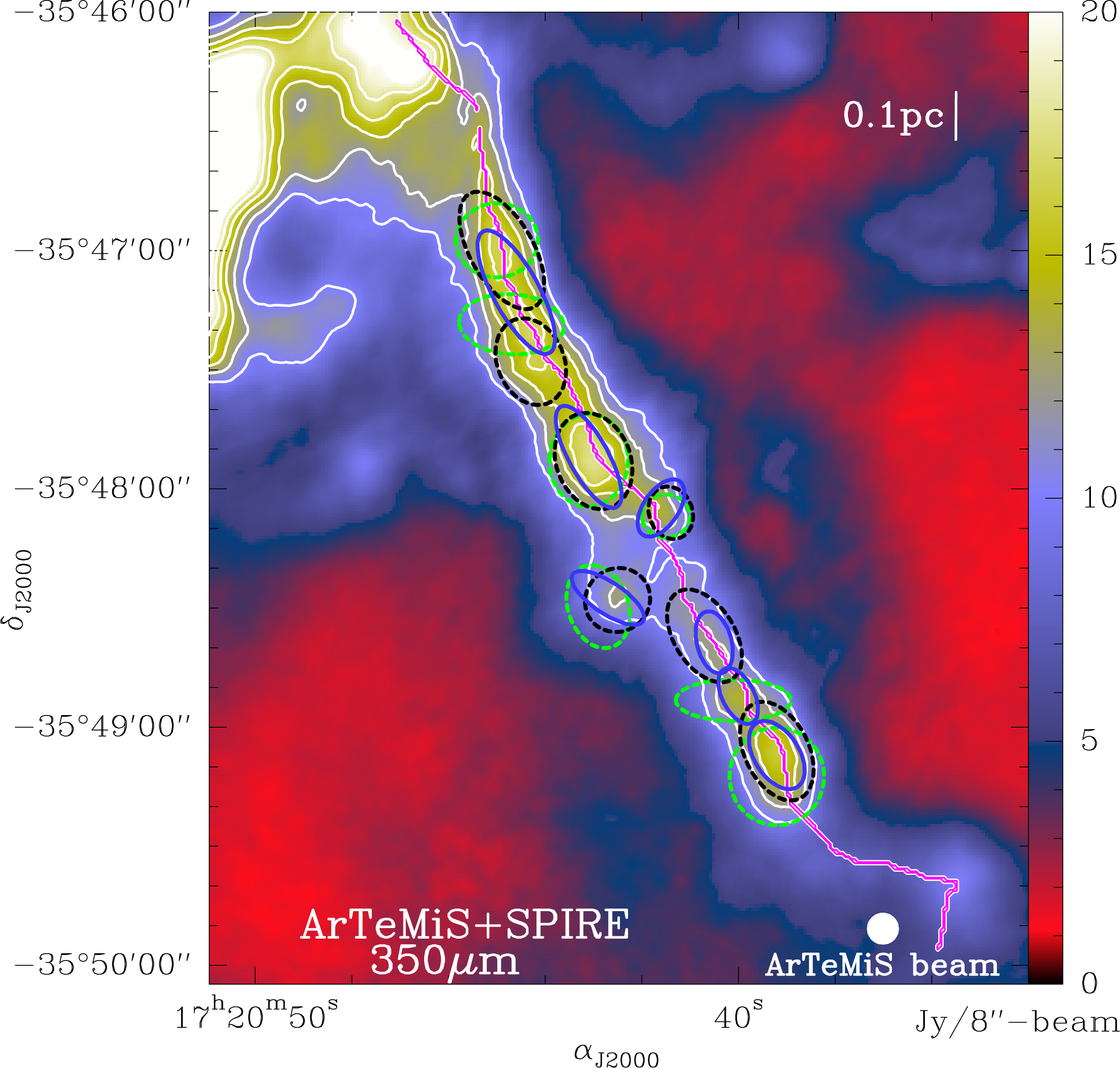}
\caption{Comparison of the clumps identified in the ArT$\acute{\rm e}$MiS 350\,$\mu$m map (color scale)
with GAUSSCLUMPS (black dashed ellipses), \textsl{getsources} (blue solid ellipses), and REINHOLD (green dashed ellipses).
}
\label{figs:comp_gaussclump_reinhold}
\end{figure}

As described in Sect.~\ref{two_levels}, we applied GAUSSCLUMPS to identify seven clumps in the ArT$\acute{\rm e}$MiS 350$\mu$m map. 
To investigate the robustness of this identification, we also applied the \textsl{getsources} algorithm, 
already used to identify compact sources in the ALMA 3.1 mm continuum map (see Sect.~\ref{sect:getsources}), 
as well as the REINOLD algorithm, available as a python package ({\it pycupid}; \citealp{Berry07}). 
REINOLD 
marks the edges of emission clumps that have shell or ring shapes. 
Then, all pixels within each ring or shell are assigned to be a single clump. Here, we adopted the following  REINOLD
parameters: RMS = 0.08 Jy beam$^{-1}$, FLATSLOPE = 1$\sigma$, FWHMBEAM = beam size, MAXBAD= 0.05, 
MINLEN = beam size/pixel size, THRESH=10, and MINPIX = beam size. 
Seven clumps were identified in the main filament. 
Figure~\ref{figs:comp_gaussclump_reinhold} 
compares the distributions of the clumps identified with GAUSSCLUMPS, \textsl{getsources}, and REINHOLD. 
The \textsl{getsources} algorithm identified seven sources within the main filament. 
Five of these seven \textsl{getsources} objects coincide with a clump found with GAUSSCLUMPS. 
REINHOLD also identified seven clumps within the main filament. One clump identified with GAUSSCLUMPS is not identified 
with REINHOLD, but is identified with \textsl{getsources}. 
All clumps identified with GAUSSCLUMPS are also identified with  \textsl{getsources}  and/or REINHOLD. 
Thus, we conclude that the seven clumps identified with GAUSSCLUMPS are robust.

\end{document}